\documentclass[preprint,12pt]{elsarticle}

\usepackage{amssymb}
\usepackage{xcolor}
\usepackage{amsmath}
\usepackage{siunitx}
\usepackage{xfrac}
\usepackage[a4paper, total={7in, 8in}]{geometry}
\usepackage[scr=boondox]{mathalfa}
\usepackage{booktabs}
\newtheorem{remark}{\bf Remark}[section]
\usepackage{float}
\usepackage{graphicx,subcaption} 
\newcounter{casecounter}

\newtheorem{case}{Case}[section]

\journal{Journal of Computational Physics}

\begin{document}

\begin{frontmatter}

\title{Centralized Gradient-Based Reconstruction for Wall Modelled Large Eddy Simulations of Hypersonic Boundary Layer Transition}

\author[label1]{Natan Hoffmann}

\author[label1,label2]{Amareshwara Sainadh Chamarthi}

\author[label1]{Steven H. Frankel}

\address[label1]{Faculty of Mechanical Engineering, Technion--Israel Institute of Technology,Haifa,3200003, Israel.}
\address[label2]{Division of Engineering and Applied Science, California Institute of Technology,Pasadena, CA,91125,USA}

\begin{abstract}

In this study, we introduce a robust central Gradient-Based Reconstruction (GBR) scheme for the compressible Navier-Stokes equations. The method leverages transformation to characteristic space, allowing selective treatment of waves from the compressible Euler equations. By averaging left- and right-biased state interpolations, a central scheme is achieved for all but the acoustic waves, which require upwinding for stability. Distinct differences were observed between transformations using either primitive or conservative variables. We evaluated the method's robustness and superiority using benchmark problems, including the two-dimensional shock entropy problem, two-dimensional viscous shock tube, and three-dimensional inviscid Taylor-Green vortex. Subsequently, we assessed the method in the context of Wall Modelled Large Eddy Simulations (WMLES), where coarse grids are used to reduce computational cost but also introduce substantial numerical dissipation. Using WMLES, we simulated oblique shock impingement on a Mach 6 disturbed boundary layer and a Mach 7.7 flow over a $15^{\circ}$ compression ramp. Our findings reveal that: 1) transformation to characteristic space using conservative variables leads to more accurate results; 2) minimizing numerical dissipation through centralized interpolation is crucial. In the compression ramp case, boundary layer separation was shifted slightly upstream, and there was an over-prediction of wall heating, likely attributable to the equilibrium-assuming wall model. Overall, this work showcases the method's potential in accurately capturing complex flow dynamics with reduced numerical dissipation.

\end{abstract}

%%Graphical abstract
% \begin{graphicalabstract}
%\includegraphics{grabs}
% \end{graphicalabstract}

%%Research highlights
% \begin{highlights}

%     \item Proposed a robust method for achieving a low-dissipation and freestream-preserving centralized Gradient-Based Reconstruction approach
%     \item Demonstrated improved accuracy with characteristic transformation using conservative variables
%     \item Applied the method in the context of Wall Modelled Large Eddy Simulation of hypersonic boundary layer transition

% \end{highlights}

\begin{keyword}

    Low Dissipation \sep  Central Scheme \sep Gradient-Based Reconstruction \sep Hypersonic \sep Wall Modelled Large Eddy Simulation

\end{keyword}

\end{frontmatter}

%% \linenumbers

%% main text
\section{Introduction} \label{sec:introduction}

% compressible turbulent flows, shocks and turbulence difficult
% gets harder with increasing reynolds number because we cant resolve all scales, so we go towards wmles
% what is wmles? it works by getting info from outer les at exchange location. thus the input is very important. need to make sure it is not riddled with unphysical dissipation
% so basically we need sensors and/or filtering to ensure that we are not adding dissipation everywhere and only appropriately
% lets take hypersonic boundary layer transition as the practical flow example and assume a high enough reynolds number where dns or wrles is not possible. heres the literature..

Numerical simulations of compressible turbulent flows present a unique set of challenges. The crux of the problem lies in the concurrent need to accurately resolve turbulence without introducing excessive numerical dissipation and to sharply capture large gradients arising from discontinuities without inducing spurious oscillations. On the one hand, regions dominated by turbulence require minimal dissipation to preserve flow details. On the other hand, areas near discontinuities demand a certain degree of dissipation to ensure solution stability. Such complexities are evident in practical flow scenarios like hypersonic boundary layer transition, supersonic jets, and scramjet combustion.

The challenge of accurately simulating compressible turbulent flows intensifies with increasing Reynolds numbers. In real-world flow scenarios, we often encounter such high Reynolds numbers, where Direct Numerical Simulation (DNS) and Wall Resolved Large Eddy Simulations (WRLES) rapidly become computationally prohibitive. This limitation stems from the need to resolve all turbulent scales, which scale with the Reynolds number. This fact is well-documented and scaling arguments are given in Yang and Griffin \cite{yang2021grid}, and Choi and Moin \cite{choi2012grid}. Given these constraints, WMLES emerges as a go-to approach for such Reynolds numbers, ensuring simulations have manageable (and/or possible) turnaround times. 

In WMLES, all inner layer dynamics are unresolved \cite{larsson2016large}. This is achieved by purposely employing a computational grid that, at a minimum, captures length scales associated with the turbulent boundary layer thickness rather than the Kolmogorov scales. The deliberate use of a coarse grid imposes an implicit spatial filter on the simulation, massively reducing computational cost from a nearly cubic relationship with Reynolds number to an almost linear scaling. Using a wall stress model, certain assumptions are made regarding the inner boundary layer such that all unresolved dynamics may be represented by a wall modelled wall shear stress and wall heat flux. The primary interaction between the LES solver and the wall model hinges on boundary conditions taken from an exchange location. This data exchange is crucial; the wall model's ability to deliver accurate wall stresses heavily depends on this input. Therefore, minimizing unphysical numerical dissipation in the resolved regions of flow is critical, rendering the choice of the LES spatial discretization method imperative.

Given this backdrop, discontinuity sensors or filtering methods are essential. These tools allow for the development of a hybrid numerical method where a discontinuity sensor identifies discontinuities and then applies either a low-dissipation or a sufficient-dissipation scheme as appropriate. This approach warrants three critical decisions: the choice of the smooth flow scheme, the discontinuity-capturing scheme, and the design of the discontinuity detector itself. The effects of these choices in the context of WMLES were recently highlighted in De Vanna et al. \cite{de2022effect} -- the choice of convective scheme had a significant effect on not only flow dynamics, but also wall stresses, which were all approximated by the same wall model.

Taking WMLES of hypersonic boundary layer transition as an example, hybrid schemes and discontinuity sensors have been used successfully in the past. Yang et al. \cite{yang2018aerodynamic} performed WMLES of the experimental/DNS work of Sandham et al. \cite{sandham2014transitional}; employing a fourth-order central scheme for flux reconstruction in smooth flow regions and an essentially non-oscillatory (ENO) scheme in regions of shocks, with the Ducros shock sensor \cite{ducros1999large} to choose between the two. Mettu and Subbareddy \cite{mettu2018wall,mettu2022wall} performed WMLES of the same case, employing a split flux approach, which used a fourth-order kinetic energy preserving scheme \cite{subbareddy2009fully} for the non-dissipative flux and a modified Steger-Warming scheme \cite{maccormack1989solution} for the dissipative flux. The dissipative flux was pre-multiplied by a spatially varying multiplicative factor governed by the Ducros shock sensor. Ganju et al. \cite{ganju2021progress} and van Noordt et al. \cite{van2022immersed} carried out WMLES of the same case using the CHAMPS code, which employs the immersed boundary method to represent geometries. The CHAMPS code also employs a split flux approach, using a different fourth-order kinetic energy and entropy preserving scheme \cite{kuya2018kinetic} for the non-dissipative flux and a fifth-order Weighted ENO (WENO) scheme for the dissipative flux. Similar to Mettu and Subbareddy \cite{mettu2018wall}, the multiplicative factor for the dissipative flux was governed by the Ducros shock sensor. Fu et al. \cite{fu2021shock} carried out WMLES of a slightly different flow configuration, in which freestream disturbances were not employed to cause boundary layer transition and instead the shock impingement angle was varied to bring about transition to turbulence. In their study, the charLES solver was used, however with a slightly different numerical method than Yang et al. \cite{yang2018aerodynamic}. One major disadvantage of using a hybrid kinetic energy/entropy preserving-ENO/WENO approach is that it is not freestream preserving \cite{ghate2023finite}. Conversely, the GBR approach was developed with freestream preservation in mind \cite{chandravamsi2023application}. In the context of hypersonic boundary layer transition simulations, this is very important to note, as unintentional freestream disturbances may significantly alter flow dynamics.

It is very important to note here that the Ducros shock sensor is dependent on a threshold value/cutoff, which has been (and is presently) treated as a parameter to the system. The cutoff level is essentially a proxy for the amount of numerical dissipation admitted to the flow field, meaning that it can adversely affect the accuracy of a given simulation. van Noordt et al. \cite{van2022immersed} and De Vanna et al. \cite{de2022effect} explored this parameter space in their studies. Ideally, the cutoff would be flow-dependent, but to the authors' knowledge, this has not been explored yet. Moreover, it is important to note that there are several variations of the original Ducros sensor in the literature. Ducros et al. \cite{ducros1999large} multiplied their sensor by the Jameson-Schmidt-Turkel artificial flux \cite{jameson1981numerical}, which is based on pressure. Some authors in recent literature do not multiply the Ducros sensor by the Jameson-Schmidt-Turkel term. Furthermore, variations to the Ducros sensor have been made to address specific issues (\cite{pirozzoli2011numerical,hendrickson2018improved}). Therefore, the Ducros cutoff parameter is far from universal.

In this study, we employ the GBR approach for inviscid flux spatial discretization. This method -- introduced by Chamarthi \cite{chamarthi2023gradient} -- employs Legendre polynomials, high-order finite differences, and monotonicity-preserving (MP) limiting; serving as a self-contained hybrid numerical method. In addition, the method proved to be computationally efficient given \textcolor{black}{its} reliance on high-order gradients and thus their availability and re-use in other parts of the solver. While the method was shown to compare well against state-of-the-art schemes such as the Targeted ENO (TENO) family of schemes introduced by Fu et al. \cite{fu2017targeted}, the MP limiting criterion employed proved to be too dissipative in later investigations. Subsequently, the Ducros shock sensor was employed in conjunction with the GBR approach, although differently than the above-mentioned studies that utilize this shock sensor. In Chamarthi et al. \cite{chamarthi2023wave}, the Ducros shock sensor was used in a broader attempt to better distinguish and treat discontinuities that may arise in compressible flows. The method relies upon characteristic transformation, which is a common step for compressible flow solvers. As a result of this transformation, each characteristic wave may be distinctly treated, providing for a more targeted approach for limiting. The results of this approach displayed a significant advantage over the base method of Chamarthi \cite{chamarthi2023gradient}. However, the method's reliance on global upwinding still proved to be a pitfall, precluding the method to be applied to simulations that are extremely sensitive to numerical dissipation. As such, in this work, we attempt to address this; introducing a robust method for using a centralized interpolation in the context of GBR and the wave appropriate discontinuity sensor approach. Broadly, we attempt to study two matters:

\begin{enumerate}
    \item If we average (centralize) the left- and right-biased interpolations in characteristic space for all but the acoustic characteristic waves, does solution fidelity improve? Are practical flow simulations of hypersonic boundary layer transition possible?
    \item Characteristic transformation can be done starting with primitive or conservative variables. Is there a difference in results when transforming from one or the other?
\end{enumerate}

The rest of the paper is organized as follows. In Sec. \ref{sec:governingEquations} we present the compressible Navier Stokes equations in curvilinear coordinates. In Sec. \ref{sec:numericalMethods}, the numerical methods used are delineated. In Sec. \ref{sec:resultsAndDiscussion}, we discuss the results of the considered test cases. Finally, in Sec. \ref{sec:conclusion}, concluding remarks are made and future work is set forth. 

\section{Governing Equations} \label{sec:governingEquations}

In this study, the three-dimensional compressible Navier-Stokes equations in conservative form were solved in curvilinear coordinates:

\begin{equation} \label{eqn:cns}
    \frac{\partial \mathbf{U}}{\partial t} + \frac{\partial \mathbf{F}^c}{\partial \xi} + \frac{\partial \mathbf{G}^c}{\partial \eta} + \frac{\partial \mathbf{H}^c}{\partial \zeta} + \frac{\partial \mathbf{F}^v}{\partial \xi} + \frac{\partial \mathbf{G}^v}{\partial \eta} + \frac{\partial \mathbf{H}^v}{\partial \zeta} = 0,
\end{equation}

\noindent where $t$ is time and $(\xi,\eta,\zeta)$ are the computational coordinates. $\mathbf{U}$ is the conserved variable vector, and $\mathbf{F}^c$, $\mathbf{G}^c$, and $\mathbf{H}^c$ are the convective flux vectors defined as:

\begin{subequations}
    \begin{gather}
        \mathbf{U} = \frac{1}{J} \begin{pmatrix}
        \rho \\
        \rho u \\
        \rho v \\
        \rho w \\
        \rho E
        \end{pmatrix},
        \mathbf{F}^c = \begin{pmatrix}
        \rho U \\
        \rho U u + \widetilde{\xi}_x p \\
        \rho U v + \widetilde{\xi}_y p \\
        \rho U w + \widetilde{\xi}_z p \\
        \rho U H 
        \end{pmatrix},
        \mathbf{G}^c = \begin{pmatrix}
        \rho V \\
        \rho V u + \widetilde{\eta}_x p \\
        \rho V v + \widetilde{\eta}_y p \\
        \rho V w + \widetilde{\eta}_z p \\
        \rho V H
        \end{pmatrix},
        \mathbf{H}^c = \begin{pmatrix}
        \rho W \\
        \rho W u + \widetilde{\zeta}_x p \\
        \rho W v + \widetilde{\zeta}_y p \\
        \rho W w + \widetilde{\zeta}_z p \\
        \rho W H
        \end{pmatrix},
        \tag{\theequation a--\theequation d}
    \end{gather}
\end{subequations}

\noindent where $J$ is the Jacobian of the transformation from physical to computational space, $\widetilde{(\cdot)}$ denotes Jacobian normalized grid metrics (e.g. $\widetilde{\xi}_x = \xi_x/J$), $\rho$ is the density, and $u$, $v$, and $w$ are the velocities in the $x$, $y$, and $z$ directions, respectively. $p$ is the pressure, $E = e + \left(u^2 + v^2 + w^2 \right)/2$ is the specific total energy, and $H = E + p/\rho$ is the specific total enthalpy. The equation of state is for a calorically perfect gas so that $e = p \left[ \rho (\gamma-1) \right]^{-1}$ is the internal energy, where $\gamma = \mathrm{c_p}/\mathrm{c_v}$ is the ratio of specific heats with $\mathrm{c_p}$ as the isobaric specific heat and $\mathrm{c_v}$ as the isochoric specific heat. $U$, $V$, and $W$ are the contravariant velocities defined as:

\begin{subequations}
    \begin{gather}
        U = \widetilde{\xi}_x u + \widetilde{\xi}_y v + \widetilde{\xi}_z w, 
        \quad
        V = \widetilde{\eta}_x u + \widetilde{\eta}_y v + \widetilde{\eta}_z w, 
        \quad
        W = \widetilde{\zeta}_x u + \widetilde{\zeta}_y v + \widetilde{\zeta}_z w,
        \tag{\theequation a--\theequation c}
    \end{gather}
\end{subequations}

\noindent where the subscripted computational coordinates are the grid metrics computed in conservative form as in Nonomura et al. \cite{nonomura2010freestream}. These metrics were computed using the same numerical method applied for spatial discretization, which ensures freestream preservation \cite{visbal2002use}. $\mathbf{F}^v$, $\mathbf{G}^v$, and $\mathbf{H}^v$ are the viscous flux vectors defined as:

\begin{subequations}
    \begin{gather}
        \mathbf{F}^v = - \begin{pmatrix}
        0 \\
        \widetilde{\xi}_x \tau_{xx} + \widetilde{\xi}_y \tau_{xy} + \widetilde{\xi}_z \tau_{xz} \\
        \widetilde{\xi}_x \tau_{yx} + \widetilde{\xi}_y \tau_{yy} + \widetilde{\xi}_z \tau_{yz} \\
        \widetilde{\xi}_x \tau_{zx} + \widetilde{\xi}_y \tau_{zy} + \widetilde{\xi}_z \tau_{zz} \\
        \widetilde{\xi}_x \beta_{x} + \widetilde{\xi}_y \beta_{y} + \widetilde{\xi}_z \beta_{z}
        \end{pmatrix},
        \quad
        \mathbf{G}^v = - \begin{pmatrix}
        0 \\
        \widetilde{\eta}_x \tau_{xx} + \widetilde{\eta}_y \tau_{xy} + \widetilde{\eta}_z \tau_{xz} \\
        \widetilde{\eta}_x \tau_{yx} + \widetilde{\eta}_y \tau_{yy} + \widetilde{\eta}_z \tau_{yz} \\
        \widetilde{\eta}_x \tau_{zx} + \widetilde{\eta}_y \tau_{zy} + \widetilde{\eta}_z \tau_{zz} \\
        \widetilde{\eta}_x \beta_{x} + \widetilde{\eta}_y \beta_{y} + \widetilde{\eta}_z \beta_{z}
        \end{pmatrix},
        \tag{\theequation a--\theequation b}
        \\
        \mathbf{H}^v = - \begin{pmatrix}
        0 \\
        \widetilde{\zeta}_x \tau_{xx} + \widetilde{\zeta}_y \tau_{xy} + \widetilde{\zeta}_z \tau_{xz} \\
        \widetilde{\zeta}_x \tau_{yx} + \widetilde{\zeta}_y \tau_{yy} + \widetilde{\zeta}_z \tau_{yz} \\
        \widetilde{\zeta}_x \tau_{zx} + \widetilde{\zeta}_y \tau_{zy} + \widetilde{\zeta}_z \tau_{zz} \\
        \widetilde{\zeta}_x \beta_{x} + \widetilde{\zeta}_y \beta_{y} + \widetilde{\zeta}_z \beta_{z}
        \end{pmatrix},
        \tag{\theequation c}
    \end{gather}
\end{subequations}

\noindent where the normal stresses are defined as:

\begin{subequations}
    \begin{gather}
        \tau_{xx} = 2 \hat{\mu} \frac{\partial u}{\partial x} + \hat{\lambda} \left(\frac{\partial u}{\partial x} + \frac{\partial v}{\partial y} + \frac{\partial w}{\partial z} \right),
        \tag{\theequation a--\theequation b}
        \quad
        \tau_{yy} = 2 \hat{\mu} \frac{\partial v}{\partial y} + \hat{\lambda} \left(\frac{\partial u}{\partial x} + \frac{\partial v}{\partial y} + \frac{\partial w}{\partial z} \right), \\[10pt]
        \tau_{zz} = 2 \hat{\mu} \frac{\partial w}{\partial z} + \hat{\lambda} \left(\frac{\partial u}{\partial x} + \frac{\partial v}{\partial y} + \frac{\partial w}{\partial z} \right),
        \tag{\theequation c}
    \end{gather}
\end{subequations}

\noindent where $\hat{\mu} = \mu \mathrm{Ma}/\mathrm{Re}$ is the scaled dynamic viscosity as a result of non-dimensionalization and Stokes' hypothesis is assumed so that $\hat{\lambda} = -2 \hat{\mu}/3$. $\mathrm{Ma} = u_{\infty} \left( \gamma R_{gas} T \right)^{-1/2}$ and $\mathrm{Re} = \rho_{\infty} u_{\infty} L_{ref}/\mu_{\infty}$ are the Mach and Reynolds numbers, respectively, where $R_{gas}$ is the universal gas constant and $\left( \cdot \right)_{\infty}$ denotes a freestream value. The shear stresses are defined as:

\begin{subequations}
    \begin{gather}
        \tau_{xy} = \tau_{yx} = \hat{\mu} \left(\frac{\partial u}{\partial y} + \frac{\partial v}{\partial x} \right),
        \quad
        \tau_{yz} = \tau_{zy} = \hat{\mu} \left(\frac{\partial v}{\partial z} + \frac{\partial w}{\partial y} \right),
        \quad
        \tau_{xz} = \tau_{zx} = \hat{\mu} \left(\frac{\partial u}{\partial z} + \frac{\partial w}{\partial x} \right),
        \tag{\theequation a--\theequation c}
    \end{gather}
\end{subequations}

\noindent and the components of $\mathbf{\beta}$ are:

\begin{subequations}
    \begin{gather}
        \beta_{x} = u \tau_{xx} + v \tau_{xy} + w \tau_{xz} + \hat{\kappa} \frac{\partial T}{\partial x},
        \tag{\theequation a--\theequation b}
        \quad
        \beta_{y} = u \tau_{yx} + v \tau_{yy} + w \tau_{yz} + \hat{\kappa} \frac{\partial T}{\partial y}, \\
        \beta_{z} = u \tau_{zx} + v \tau_{zy} + w \tau_{zz} + \hat{\kappa} \frac{\partial T}{\partial z},
        \tag{\theequation c}
    \end{gather}
\end{subequations}

\noindent where $\hat{\kappa} = \hat{\mu} \left[ (\gamma-1)\mathrm{Pr} \right]^{-1}$ is the scaled thermal conductivity, $\mathrm{Pr}$ is the Prandtl number, and $T$ is the temperature. The dynamic viscosity was taken as a function of temperature by Sutherland's law:

\begin{equation}
    \mu(T) = T^{3/2} \frac{1 + S/T_{ref}}{T + S/T_{ref}},
    \label{eqn:sutherlandsLaw}
\end{equation}

\noindent where $T_{ref}$ is the reference temperature and $S = \SI{110.4}{\kelvin}$ is Sutherland's constant. The equations were non-dimensionalized using the freestream density $\rho_{\infty}$, the freestream speed of sound $c_{\infty}$, reference length $L_{ref}$, the freestream temperature $T_{\infty}$, and the freestream dynamic viscosity $\mu_{\infty}$ such that the temperature was related to pressure and density via $p = \rho T / \gamma$.

\begin{remark}
    \textcolor{black}{Technically, for the WMLES cases, we solved the Favre-averaged Navier-Stokes equations. The sub-grid scale (SGS) terms that arise in these equations were treated implicitly. Thus, the WMLES cases are implicit WMLES.}
\end{remark}

\section{Numerical Methods} \label{sec:numericalMethods}

In this work, we employed the GBR method of Chamarthi \cite{chamarthi2023gradient} tailored to effective discontinuity capturing \cite{chamarthi2023wave} by a characteristic wave appropriate sensor. In Chamarthi et al. \cite{chamarthi2023wave}, it was shown that selectively treating discontinuities incumbent to compressible flows via characteristic transformation greatly improves solution quality in a physically consistent manner. In this work, we show that not only can we selectively sense and treat the characteristic waves, but we can also use centralized interpolations for three out of five of these waves, further reducing unnecessary numerical dissipation. In what follows, we delineate the details of this method. After, we briefly present the viscous flux discretization, wall stress model used for the final test case, and the time integration method.

\subsection{Convective Flux Spatial Discretization Scheme}

Using a conservative numerical method, the governing equations cast in semi-discrete form for a curvilinear cell $I_{i,j,k} = \left[ \xi_{i-\frac{1}{2}}, \xi_{i+\frac{1}{2}} \right] \times \left[ \eta_{i-\frac{1}{2}}, \eta_{i+\frac{1}{2}} \right] \times \left[ \zeta_{i-\frac{1}{2}}, \zeta_{i+\frac{1}{2}} \right]$ can be expressed via the following ordinary differential equation: 

\begin{align}
    \begin{aligned}
        \frac{\mathrm{d}}{\mathrm{d} t} \mathbf{U}_{i,j,k} = \mathbf{Res}_{i,j,k} = &- \left. \frac{\mathrm{d} \mathbf{F}^c}{\mathrm{d} \xi} \right|_{i,j,k} - \left. \frac{\mathrm{d} \mathbf{G}^c}{\mathrm{d} \eta} \right|_{i,j,k} - \left. \frac{\mathrm{d} \mathbf{H}^c}{\mathrm{d} \zeta} \right|_{i,j,k} \\ 
        &+ \left. \frac{\mathrm{d} \mathbf{F}^v}{\mathrm{d} \xi} \right|_{i,j,k} + \left. \frac{\mathrm{d} \mathbf{G}^v}{\mathrm{d} \eta} \right|_{i,j,k} + \left. \frac{\mathrm{d} \mathbf{H}^v}{\mathrm{d} \zeta} \right|_{i,j,k},
    \end{aligned}
\end{align}

\noindent where $\mathbf{Res}_{i,j,k}$ is the residual function,and the remaining terms are cell center numerical flux derivatives of the physical fluxes in Eqn. \ref{eqn:cns}. For brevity, we continue with only the $\xi$-direction, however, the following may be extended to all three dimensions straightforwardly. Moreover, we drop the $j$ and $k$ indices in the interest of clarity. The cell center numerical convective flux derivative is expressed as:

\begin{equation}
    \left. \frac{\mathrm{d} \mathbf{F}^c}{\mathrm{d} \xi} \right|_{i} = \frac{1}{\Delta \xi} \left( \mathbf{F}^{c}_{i+\frac{1}{2}} - \mathbf{F}^{c}_{i-\frac{1}{2}} \right),
\end{equation}

\noindent where $i \pm \frac{1}{2}$ indicates right and left cell interface values, respectively. $\mathbf{F}^c_{i \pm \frac{1}{2}}$ are computed using an approximate Riemann solver, since a Riemann problem exists at each cell interface. The interface numerical convective fluxes are computed from:

\begin{equation}\label{eqn:riemannProblem}
    \mathbf{F}^c_{i \pm \frac{1}{2}} = \frac{1}{2} \left[ \mathbf{F}^c \left( \mathbf{U}^{L}_{i \pm \frac{1}{2}} \right) + \mathbf{F}^c \left( \mathbf{U}^{R}_{i \pm \frac{1}{2}} \right) \right] - \frac{1}{2} \left| \mathbf{A}_{i \pm \frac{1}{2}} \right| \left( \mathbf{U}^{R}_{i \pm \frac{1}{2}} - \mathbf{U}^{L}_{i \pm \frac{1}{2}} \right),
\end{equation}

\noindent where the $L$ and $R$ superscripts denote the left- and right-biased states, respectively, and $\left| \mathbf{A}_{i \pm \frac{1}{2}} \right|$ denotes the convective flux Jacobian. The objective is to obtain the left- and right-biased states. These were computed with the GBR method, which will be explained in the following subsection.

\subsubsection{Gradient-Based Reconstruction Method: Linear Scheme}

GBR methods employ the first two moments of the Legendre polynomial evaluated on $\xi_{i-\frac{1}{2}} \leq \xi \leq \xi_{i+\frac{1}{2}}$ for interpolation. This may be written for a general variable, $\phi$, as:

\begin{equation}\label{eqn:legendre}
    \textcolor{black}{\phi(\xi) = \phi_{i} + \phi'_{i} (\xi-\xi_i) + \frac{3 \phi''_{i}}{2}  \mathscr{K} \left[ (\xi-\xi_i)^{2} - \frac{\Delta \xi^{2}_{i}}{12} \right],}
\end{equation}

\noindent where $\phi'_{i}$ and $\phi''_{i}$ respectively represent the first and second derivatives of $\phi_{i}$. If $\xi = \xi_i + \Delta \xi/2$ and $\mathscr{K} = 1/3$, the following equations for the left- and right-biased states are obtained:
    
\begin{subequations}
    \begin{gather}
        \textcolor{black}{\phi^{L, Linear}_{i+\frac{1}{2}} = \phi_{i} + \frac{\Delta \xi}{2} \phi'_{i} + \frac{\Delta \xi^2}{12} \phi''_{i},}
        \quad
        \textcolor{black}{\phi^{R, Linear}_{i+\frac{1}{2}} = \phi_{i+1} - \frac{\Delta \xi}{2} \phi'_{i+1} + \frac{\Delta \xi^2}{12} \phi''_{i+1}.}
        \tag{\theequation a--\theequation b}
    \end{gather}
    \label{eqn:legendreInterpolation}
\end{subequations}

\noindent In this work, $\phi'_{i}$ was computed using eighth order explicit central differences:
    
\begin{equation}
    \phi'_{i} = \frac{1}{\Delta \xi} \left( \frac{1}{280} \phi_{i-4} - \frac{4}{105} \phi_{i-3} + \frac{1}{5} \phi_{i-2} - \frac{4}{5} \phi_{i-1} + \frac{4}{5} \phi_{i+1} - \frac{1}{5} \phi_{i+2} +\frac{4}{105} \phi_{i+3} - \frac{1}{280} \phi_{i+4} \right).
    \label{eqn:firstDerivative}
\end{equation}

\noindent $\phi''_{i}$ was computed from:

\begin{equation}
    \phi''_{i} = \frac{2}{\Delta \xi^2} \left( \phi_{i+1} - 2 \phi_{i} + \phi_{i-1} \right) - \frac{1}{2 \Delta \xi} \left( \phi'_{i+1} - \phi'_{i-1} \right).
    \label{eqn:secondDerivative}
\end{equation}

\noindent Since $\phi$ is an arbitrary variable, either primitive ($\mathbf{P}$), conservative ($\mathbf{U}$), or characteristic ($\mathbf{C}$) variables may be used.

\subsubsection{Gradient-Based Reconstruction Method: Non-Linear Scheme}

Eqns. \ref{eqn:legendreInterpolation} are linear interpolations. Therefore, they may be susceptible to oscillations in the presence of discontinuities. So, MP limiting was employed. The following delineates the MP limiting procedure for the left-biased state, however, the procedure is the same for the right-biased state. The MP limiting criterion is:

\begin{equation}
    \mathbf{U}^{L}_{i+\frac{1}{2}} = 
    \begin{cases}
        \mathbf{U}^{L,Linear}_{i+\frac{1}{2}} & \text{if } \left( \mathbf{U}^{L,Linear}_{i+\frac{1}{2}} - \mathbf{U}_i \right) \left( \mathbf{U}^{L,Linear}_{i+\frac{1}{2}} - \mathbf{U}^{L,MP}_{i+\frac{1}{2}} \right) \leq 10^{-20}, \\[5pt]
        \mathbf{U}^{L,Non-Linear}_{i+\frac{1}{2}} & \text{otherwise},
    \end{cases}
    \label{eqn:mpLimitingCriterion}
\end{equation}

\noindent where $\mathbf{U}^{L,Linear}_{i+\frac{1}{2}}$ corresponds to Eqn. \ref{eqn:legendreInterpolation}a, and the remaining terms are:

\begin{subequations}\label{eqn:mp_improve}
    \begin{alignat}{2}
        &\mathbf{U}^{L,Non-Linear}_{i+\frac{1}{2}} &&= \mathbf{U}^{L,Linear}_{i+\frac{1}{2}} + \text{minmod} \left( \mathbf{U}^{L,MIN}_{i+\frac{1}{2}} - \mathbf{U}^{L,Linear}_{i+\frac{1}{2}}, \mathbf{U}^{L,MAX}_{i+\frac{1}{2}} - \mathbf{U}^{L,Linear}_{i+\frac{1}{2}} \right),
        \\[5pt]
        &\mathbf{U}^{L,MP}_{i+\frac{1}{2}} &&= \mathbf{U}^{L,Linear}_{i+\frac{1}{2}} + \text{minmod} \left[ \mathbf{U}_{i+1}-\mathbf{U}_{i}, \mathscr{A} \left( \mathbf{U}_{i}-\mathbf{U}_{i-1} \right) \right], 
        \\[5pt]
        &\mathbf{U}^{L,MIN}_{i+\frac{1}{2}} &&= \max \left[ \min \left( \mathbf{U}_{i}, \mathbf{U}_{i+1}, \mathbf{U}^{L,MD}_{i+\frac{1}{2}} \right), \min \left( \mathbf{U}_{i}, \mathbf{U}^{L,UL}_{i+\frac{1}{2}}, \mathbf{U}^{L,LC}_{i+\frac{1}{2}} \right) \right],
        \\[5pt]
        &\mathbf{U}^{L,MAX}_{i+\frac{1}{2}} &&= \min \left[ \max \left( \mathbf{U}_{i}, \mathbf{U}_{i+1}, \mathbf{U}^{L,MD}_{i+\frac{1}{2}} \right), \max \left( \mathbf{U}_{i}, \mathbf{U}^{L,UL}_{i+\frac{1}{2}}, \mathbf{U}^{L,LC}_{i+\frac{1}{2}} \right) \right],
        \\[5pt]
        &\mathbf{U}^{L,MD}_{i+\frac{1}{2}} &&= \frac{1}{2} \left( \mathbf{U}_{i} + \mathbf{U}_{i+1} \right) - \frac{1}{2} d^{L,M}_{i+\frac{1}{2}},
        \\[5pt]
        &\mathbf{U}^{L,UL}_{i+\frac{1}{2}} &&= \mathbf{U}_{i} + 4 \left( \mathbf{U}_{i} - \mathbf{U}_{i-1} \right),
        \\[5pt]
        &\mathbf{U}^{L,LC}_{i+\frac{1}{2}} &&= \frac{1}{2} \left( 3 \mathbf{U}_{i} - \mathbf{U}_{i-1} \right) + \frac{4}{3} d^{L,M}_{i-\frac{1}{2}},
        \\[5pt]
        &d^{L,M}_{i+\frac{1}{2}} &&= \text{minmod} \left( d_i, d_{i+1} \right),
        \\[5pt]
        &d_i &&= 2 \left( \mathbf{U}_{i+1} - 2\mathbf{U}_{i} + \mathbf{U}_{i-1} \right) - \frac{\Delta x}{2} \left( \mathbf{U}'_{i+1} - \mathbf{U}'_{i-1} \right),
    \end{alignat}
\end{subequations}

\noindent where $\mathscr{A} = 4$ and $\text{minmod} \left( a,b \right) = \frac{1}{2} \left[ \text{sgn}(a) + \text{sgn}(b) \right] \min \left( \left| a \right|, \left| b \right| \right)$. \textcolor{black}{The GBR method that employs explicit finite differences and the above MP limiter is called MEG (\textbf{M}onotonicity Preserving \textbf{E}xplicit \textbf{G}radient). For a more detailed description and schematic representation of the MP limiter, readers may refer to the original work of Suresh and Hyunh \cite{suresh1997accurate}.}

\subsubsection{Ducros Shock Sensor}

While MP limiting effectively mitigates oscillations arising from discontinuities, the detection algorithm in Eqn. \ref{eqn:mpLimitingCriterion} can become too sensitive and cause excessive dissipation. To remedy this issue, the Ducros shock sensor, which was designed specifically to sense shocks, can be used \cite{fang2013optimized}:

\begin{equation}
    \Omega^{d}_{i,j,k} = \theta^{d}_{i,j,k} \frac{ \left( \nabla \cdot \mathbf{u} \right)^2}{ \left( \nabla \cdot \mathbf{u} \right)^2 + \left| \nabla \times \mathbf{u} \right|^2 + \epsilon}, \quad \text{for } d \in \{ \xi, \eta, \zeta \}
    \label{eqn:ducros}
\end{equation}

\noindent where $\epsilon = 10^{-40}$ to avoid division by zero and,

\begin{subequations}
    \begin{gather}
        \theta^{\xi}_{i,j,k} = \frac{ \left| p_{i+1}-2 p_{i}+p_{i-1} \right| }{ \left| p_{i+1}+2 p_{i}+p_{i-1} \right| },
        \quad
        \theta^{\eta}_{i,j,k} = \frac{ \left| p_{j+1}-2 p_{j}+p_{j-1} \right| }{ \left| p_{j+1}+2 p_{j}+p_{j-1} \right| },
        \quad
        \theta^{\zeta}_{i,j,k} = \frac{ \left| p_{k+1}-2 p_{k}+p_{k-1} \right| }{ \left| p_{k+1}+2 p_{k}+p_{k-1} \right| },
        \tag{\theequation a--\theequation c}
    \end{gather}
\end{subequations}

\noindent and $\mathbf{u}$ is the velocity vector.  We modify $\Omega^{d}_{i,j,k}$ by using \textcolor{black}{its} maximum value in a three cell neighborhood in all directions:

\begin{equation}
    \Omega^{d}_{i,j,k} = \max \left( \Omega^{d}_{i+m,j+m,k+m} \right), \quad \text{for } m = -1,0,1. 
\end{equation}

\noindent Then, taking the $\xi$ direction as an example, Eqn. \ref{eqn:mpLimitingCriterion} is modified to:

\begin{equation}
    \mathbf{U}^{L}_{i+\frac{1}{2}} = 
    \begin{cases}
        \mathbf{U}^{L,Linear}_{i+\frac{1}{2}} & \text{if } \overline{\Omega^{\xi}_{i,j,k}} \leq 0.01, \\[5pt]
        \mathbf{U}^{L,Non-Linear}_{i+\frac{1}{2}} & \text{otherwise}.
    \end{cases}
    \label{eqn:ducrosLimitingCriterion}
\end{equation}

\subsubsection{Wave Appropriate Discontinuity Sensor}

For coupled hyperbolic equations like the Euler equations, shock-capturing should be carried out using characteristic variables for \textit{cleanest} results \cite{van2006upwind}. The wave appropriate discontinuity sensor algorithm takes advantage of the transformation from physical to characteristic space:

\begin{enumerate}
    
    \item Compute Roe-averaged variables following Blazek \cite{blazek2015computational} (Equation 4.89) to construct the left, $\mathbf{L}_n$, and right, $\mathbf{R}_n$, eigenvectors of the normal convective flux Jacobian. \\
    
    \item Transform $\mathbf{U}_{i}$, $\mathbf{U}'_{i}$, and $\mathbf{U}''_{i}$ to characteristic space by multiplying them by $\mathbf{L}_n$:

    \begin{subequations}
        \begin{gather}
            \mathbf{C}_{i+m,b} = \mathbf{L}_{n,i+\frac{1}{2}} \mathbf{U}_{i+m}, 
            \quad
            \mathbf{C}'_{i+m,b} = \mathbf{L}_{n,i+\frac{1}{2}} \mathbf{U}'_{i+m},
            \quad
            \mathbf{C}''_{i+m,b} = \mathbf{L}_{n,i+\frac{1}{2}} \mathbf{U}''_{i+m}, 
            \tag{\theequation a--\theequation c}
        \end{gather}
    \end{subequations}

    for $m = -2,-1,0,1,2,3$ and $b = 1,2,3,4,5$, representing the vector of characteristic variables. In our implementation: 

    \begin{table}[H]
        \centering
        \caption{Characteristic wave types.}
        \begin{tabular}{c c c}
            \hline
            \hline
            $b = 1,5$ & $b = 2$ & $b = 3,4$ \\
            \hline
            Acoustic & Entropy/Contact & Shear/Vortical  \\
            \hline
            \hline
        \end{tabular}
        \label{tab:characteristicWaveStructure}
    \end{table}
        
    The second characteristic variable, $\mathbf{C}_{i+m,2}$, corresponds to what is known in one-dimension as the entropy or contact wave. $\mathbf{C}_{i+m,2}$ requires limiting in the presence of contact discontinuities, which significantly improves solution quality in a physically consistent manner. \\

    \item Using Eqns. \ref{eqn:legendreInterpolation}, obtain the unlimited interpolation to cell interfaces in characteristic space via:

\begin{subequations}
    \begin{gather}
        \textcolor{black}{\mathbf{C}^{L, Linear}_{i+\frac{1}{2},b} = \mathbf{C}_{i,b} + \frac{\Delta \xi}{2} \mathbf{C}'_{i,b} + \frac{\Delta \xi^2}{12} \mathbf{C}''_{i,b},
        \quad
        \mathbf{C}^{R, Linear}_{i+\frac{1}{2},b} = \mathbf{C}_{i+1,b} - \frac{\Delta \xi}{2} \mathbf{C}'_{i+1,b} + \frac{\Delta \xi^2}{12} \mathbf{C}''_{i+1,b}.}
        \tag{\theequation a--\theequation b}
    \end{gather}
    \label{eqn:unlimitedCharacteristicInterpolation}
\end{subequations}

\noindent The left-biased interpolation is then treated by the following limiting algorithm:

\begin{equation}
    \mathbf{C}^{L}_{i+\frac{1}{2},b} = 
    \begin{cases}
        \mathbf{C}^{L,Non-Linear}_{i+\frac{1}{2},b} & \text{if } b = 2 \text{ and } \left( \mathbf{C}^{L,Linear}_{i+\frac{1}{2}} - \mathbf{C}_i \right) \left( \mathbf{C}^{L,Linear}_{i+\frac{1}{2}} - \mathbf{C}^{L,MP}_{i+\frac{1}{2}} \right) \leq 10^{-20}, 
        \\[10pt]
        \mathbf{C}^{L,Non-Linear}_{i+\frac{1}{2},b} & \text{if } b \neq 2 \text{ and } \overline{\Omega^d_{i,j,k}} > 0.01,
        \\[10pt]
        \mathbf{C}^{L,Linear}_{i+\frac{1}{2},b} & \text{otherwise},
    \end{cases}
    \label{eqn:newSensorCriterion}
\end{equation}

\noindent and likewise for the right-biased interpolation.

\item After obtaining $\mathbf{C}^{L,R}_{i+\frac{1}{2},b}$, the reconstructed states are then recovered by projecting the characteristic variables back to physical fields:

\begin{equation}\label{eqn:characteristicToPhysical}
    \mathbf{U}^{L,R}_{i+\frac{1}{2}} = \mathbf{R}_{n,i+\frac{1}{2}} \mathbf{C}^{L,R}_{i+\frac{1}{2}}.
\end{equation}

\end{enumerate}

This scheme is denoted as MEG-S (selective) as in Chamarthi et al. \cite{chamarthi2023wave}.

\subsubsection{Wave Appropriate Centralization}

In Chamarthi et al. \cite{chamarthi2023wave} two issues were addressed: 

\begin{itemize}
    \item reducing unnecessary numerical dissipation due to the application of the MP limiter for all characteristic waves by employing the Ducros shock sensor for all but the entropy/contact characteristic wave,
    \item and maintaining MP limiting for the entropy/contact characteristic wave to reduce spurious oscillations due to the inappropriate application of the Ducros shock sensor.
\end{itemize}

\noindent The final scheme, denoted MEG-S, however, solely relied on an upwinded linear interpolation (Eqns. \ref{eqn:legendreInterpolation}). Thus, even without the selective limiting of this interpolation in characteristic space, excess numerical dissipation was unavoidable. Certain physical phenomena may be extremely sensitive to additional unphysical dissipation. As such, given the application of GBR to numerical simulations of such physical phenomena (\cite{hoffmann2023large}, \cite{chandravamsi2023application}), this shortcoming needed to be addressed. \\

Upon investigation, it was found that using a centralized interpolation for all but the characteristic acoustic waves was a robust and superior solution. To completely minimize dissipation from the convective spatial discretization scheme, one can average the left- and right-biased interpolations to effectively achieve a central scheme:

\begin{subequations}\label{eqn:centralScheme}
    \begin{gather}
        \phi^{L}_{i+\frac{1}{2}} = \phi^{C}_{i+\frac{1}{2}} = \left( 1 - A \right) \phi^{L, Linear}_{i+\frac{1}{2}} + A \phi^{R, Linear}_{i+\frac{1}{2}},
        \tag{\theequation a}
        \\[5pt]
        \phi^{R}_{i+\frac{1}{2}} = \phi^{C}_{i+\frac{1}{2}} = B \phi^{L, Linear}_{i+\frac{1}{2}} + \left( 1 - B \right) \phi^{R, Linear}_{i+\frac{1}{2}},
        \tag{\theequation b}
    \end{gather}
\end{subequations}

\noindent where $A = B = 0.5$ and $\left( \cdot \right)^C$ denotes the centralized interpolation. This effectively causes the second term on the right-hand-side of Eqn. \ref{eqn:riemannProblem} to reduce to zero. In effect, Eqn. \ref{eqn:newSensorCriterion} becomes:

\begin{equation}
    \mathbf{C}^{L}_{i+\frac{1}{2},b} = 
    \left\{
    \begin{array}{ll}
        \text{if } b = 1,5\text{:} & \begin{cases}
            \mathbf{C}^{L,Non-Linear}_{i+\frac{1}{2},b} & \text{if } \overline{\Omega^d_{i,j,k}} > 0.01, 
            \\[10pt]
            \mathbf{C}^{L,Linear}_{i+\frac{1}{2},b} & \text{otherwise},
        \end{cases} 
        \\[30pt]
        \text{if } b = 2\text{:} & \begin{cases}
            \mathbf{C}^{C,Non-Linear}_{i+\frac{1}{2},b} & \text{if } \left( \mathbf{C}^{C,Linear}_{i+\frac{1}{2}} - \mathbf{C}_i \right) \left( \mathbf{C}^{C,Linear}_{i+\frac{1}{2}} - \mathbf{C}^{C,MP}_{i+\frac{1}{2}} \right) \leq 10^{-20}, 
            \\[10pt]
            \mathbf{C}^{C,Linear}_{i+\frac{1}{2},b} & \text{otherwise},
        \end{cases} 
        \\[30pt]
        \text{if } b = 3,4\text{:} & \begin{cases}
            \mathbf{C}^{C,Non-Linear}_{i+\frac{1}{2},b} & \text{if } \overline{\Omega^d_{i,j,k}} > 0.01, 
            \\[10pt]
            \mathbf{C}^{C,Linear}_{i+\frac{1}{2},b} & \text{otherwise}.
        \end{cases}
    \end{array}
    \right.
    \label{eqn:centralizedWaveSensorCriterion}
\end{equation}

\noindent Note that for the acoustic waves, the upwinded interpolation is used for the linear interpolation and the evaluation of the non-linear interpolation. In contrast, for the remaining waves, the centralized interpolation is used for the linear interpolation and even in the evaluation of the non-linear interpolation. This is important to recognize, as we found that the use of the central scheme to evaluate the non-linear interpolation (for $b = 2,3,4$) improved solution quality, as well. This resultant scheme is denoted MEG-C (centralized).

\subsubsection{Approximate Riemann Solver} \label{section:approximateRiemannSolver}

In this Section, we present the approximate Riemann solvers used. We used a hybrid HLL-HLLC approximate Riemann solver for the WMLES cases. The remaining cases all used the HLLC approximate Riemann solver. Starting with the HLLC approximate Riemann solver in generalized coordinates, only the $\xi$ direction is presented here. This flux is defined as:

\begin{equation}
    \mathbf{F}^{HLLC}_{i \pm \frac{1}{2}} = 
    \begin{cases}
        \mathbf{F}^{L}, & \text{if } 0 \leq S^L, \\
        \mathbf{F}^{L}_{*}, & \text{if } S^L \leq 0 \leq S_{*}, \\
        \mathbf{F}^{R}_{*}, & \text{if } S_{*} \leq 0 \leq S^{R}, \\
        \mathbf{F}^{R}, & \text{if } 0 \geq S^{R},
    \end{cases}
\end{equation}

\noindent where,

\begin{equation}
    \mathbf{F}^{K}_{*} = \mathbf{F}^{K} + S^K \left(\mathbf{U}^{K}_{*} - \mathbf{U}^K \right),
\end{equation}

\noindent where $K$ denotes the left ($L$) or right ($R$) state. Extending from Pathak and Shukla \cite{pathak2016adaptive}, the star state is defined as:

\begin{equation}
    \mathbf{U}^{K}_{*} = \rho^K \left( \dfrac{S^K - U^K}{S^K - S_{*}} \right) 
    \begin{Bmatrix}
        1 
        \\[15pt]
        \dfrac{\widehat{\xi}_x S_* + \left( \widehat{\xi}^{2}_{y} + \widehat{\xi}^{2}_{z} \right) u^K - \widehat{\xi}_y \widehat{\xi}_x v^K - \widehat{\xi}_z \widehat{\xi}_x w^K}{\widehat{\xi}^{2}_{x} + \widehat{\xi}^{2}_{y} + \widehat{\xi}^{2}_{z}} 
        \\[15pt]
        \dfrac{\widehat{\xi}_y S_* - \widehat{\xi}_x \widehat{\xi}_y u^K + \left( \widehat{\xi}^{2}_{x} + \widehat{\xi}^{2}_{z} \right) v^K - \widehat{\xi}_z \widehat{\xi}_y w^K}{\widehat{\xi}^{2}_{x} + \widehat{\xi}^{2}_{y} + \widehat{\xi}^{2}_{z}} 
        \\[15pt]
        \dfrac{\widehat{\xi}_z S_* - \widehat{\xi}_x \widehat{\xi}_z u^K - \widehat{\xi}_y \widehat{\xi}_z v^K + \left(\widehat{\xi}^{2}_{x} + \widehat{\xi}^{2}_{y} \right) w^K}{\widehat{\xi}^{2}_{x} + \widehat{\xi}^{2}_{y} + \widehat{\xi}^{2}_{z}} 
        \\[15pt]
        E^K + \left( S_{*} - U^K \right) \left[ \dfrac{S_{*}}{\widehat{\xi}^{2}_{x} + \widehat{\xi}^{2}_{y} + \widehat{\xi}^{2}_{z}} + \dfrac{p^K}{\rho^K \left(S^K-U^K \right)} \right]
    \end{Bmatrix},
\end{equation}

\noindent where $\widehat{({\cdot})}$ denotes grid metrics interpolated from cell centers to cell interfaces. Note that the temporal metrics have been omitted. The left, right, and star wave speeds are respectively:

\begin{equation}
    S^L = \min \left( U^L - c^L \sqrt{\widehat{\xi}^{2}_{x} + \widehat{\xi}^{2}_{y} + \widehat{\xi}^{2}_{z}},\breve{U} - \breve{c} \right),
\end{equation}

\begin{equation}
    S^R = \max \left( U^R + c^R \sqrt{\widehat{\xi}^{2}_{x} + \widehat{\xi}^{2}_{y} + \widehat{\xi}^{2}_{z}},\breve{U} + \breve{c} \right),
\end{equation}

\begin{equation}
    S_{*} = \frac{\rho^R U^R \left( S^R - U^R \right) - \rho^L U^L \left( S^L - U^L \right) + \left( p^L - p^R \right) \left(\widehat{\xi}^{2}_{x} + \widehat{\xi}^{2}_{y} + \widehat{\xi}^{2}_{z} \right)}{\rho^R \left( S^R - U^R \right) - \rho^L \left( S^L - U^L \right)},
\end{equation}

\noindent where $\breve{({\cdot})}$ denotes Roe-averaged variables. 

The HLL flux is defined as:

\begin{equation}
    \mathbf{F}^{HLL}_{i \pm \frac{1}{2}} = 
    \begin{cases}
        \mathbf{F}^{L}, & \text{if } 0 \leq S^L, 
        \\[15pt]
        \dfrac{S^R \mathbf{F}^{L} - S^L \mathbf{F}^{R} + S^L S^R \left(\mathbf{U}^R - \mathbf{U}^L \right)}{S^L - S^R}, & \text{if } S^L \leq 0 \leq S^R, 
        \\[15pt]
        \mathbf{F}^{R}, & \text{if } 0 \geq S^{R}.
    \end{cases}
\end{equation}

\noindent As abovementioned, for the WMLES cases, a hybrid HLL-HLLC approach was employed. The hybridization criterion was:

\begin{equation}
    \mathbf{F}_{i \pm \frac{1}{2}} = 
    \begin{cases}
        \mathbf{F}^{HLL}_{i \pm \frac{1}{2}}, & \text{if } \overline{\Omega^d_{i,j,k}} > 0.01, 
        \\[15pt]
        \mathbf{F}^{HLLC}_{i \pm \frac{1}{2}}, & \text{otherwise.}
    \end{cases}
    \label{eqn:hybridRiemannSolver}
\end{equation}

\subsection{Viscous Flux Spatial Discretization Scheme}

In this subsection, we present the spatial discretization of the numerical viscous fluxes. We used the fourth-order $\alpha$-damping scheme of Chamarthi \cite{chamarthi2023gradient} and Chamarthi et al. \cite{chamarthi2022importance}, which is based on the $\alpha$-damping approach of Nishikawa \cite{nishikawa2011two}. The importance of using such a viscous flux discretization in the context of turbulent flows was recently highlighted in Chamarthi et al. \cite{chamarthi2023role}. For simplicity and without loss of generality, we present a one-dimensional scenario. The cell center numerical viscous flux derivative is:

\begin{equation}
    \left. \frac{\mathrm{d} \mathbf{F}^v}{\mathrm{d} \xi} \right|_{i} = \frac{1}{\Delta \xi} \left( \mathbf{F}^{v}_{i+\frac{1}{2}} - \mathbf{F}^{v}_{i-\frac{1}{2}} \right),
\end{equation}

\noindent The cell interface numerical viscous flux is:

\begin{equation}
    \mathbf{F}^v_{i+\frac{1}{2}} = 
    \begin{pmatrix}
        0 \\
        -\tau_{i+\frac{1}{2}} \\
        -\tau_{i+\frac{1}{2}} u_{i+\frac{1}{2}} + q_{i+\frac{1}{2}} \\
    \end{pmatrix},
\end{equation}

\noindent where,

\begin{subequations}
    \begin{gather}
        \tau_{i+\frac{1}{2}} = \frac{4}{3} \hat{\mu}_{i+\frac{1}{2}} \left. \frac{\partial u}{\partial x} \right|_{i+\frac{1}{2}},
        \quad
        q_{i+\frac{1}{2}} = -\hat{\kappa}_{i+\frac{1}{2}} \left. \frac{\partial T}{\partial x} \right|_{i+\frac{1}{2}}.
        \tag{\theequation a--\theequation b}
        \label{eqn:interfaceViscousStress}
    \end{gather}
\end{subequations}

\noindent As mentioned by Eqn. \ref{eqn:sutherlandsLaw}, the dynamic viscosity was computed using Sutherland's Law (unless otherwise stated). Since the dynamic viscosity is required at the cell interface, the cell interface temperature is used. The cell interface temperature was computed via an arithmetic average, i.e. $T_{i+\frac{1}{2}} = \left(T_i + T_{i+1} \right)/2$. For an arbitrary variable, $\phi$, the $\alpha$-damping approach computes cell interface gradients as:

\begin{equation}
    \left. \frac{\partial \phi}{\partial x} \right|_{i+ \frac{1}{2}} = \frac{1}{2} \left( \left. \frac{\partial \phi}{\partial x} \right|_{i} + \left. \frac{\partial \phi}{\partial x} \right|_{i+1} \right) + \frac{\alpha}{2 \Delta x} \left( \phi_R - \phi_L \right), 
\end{equation}

\noindent where, 

\begin{subequations}
    \begin{alignat}{2}
        &\phi_L &&= \phi_i + \left. \frac{\partial \phi}{\partial x} \right|_{i} \frac{\Delta x}{2} + \beta \left( {\phi}_{i+1} - 2 {\phi}_{i} + {\phi}_{i-1} \right),
        \\[5pt]
        &\phi_R &&= \phi_{i+1} - \left. \frac{\partial \phi}{\partial x} \right|_{i+1} \frac{\Delta x}{2} + \beta \left( {\phi}_{i+2} - 2 {\phi}_{i+1} + {\phi}_{i} \right),
    \end{alignat}
\end{subequations}

\noindent where, in this work, $\alpha = 4$ and $\beta = 0$. The gradients at cell centers were the same ones computed in Eqn. \ref{eqn:firstDerivative}.

\subsection{Wall-Stress Model}

In this work, the compressible equilibrium ODE wall model of Kawai and Larsson \cite{kawai2012wall} was used to model the unresolved, inner layer dynamics of the boundary layer for the WMLES cases. In this model, the flow is assumed to be steady, parallel, and in equilibrium so that the convective and pressure gradient terms of the compressible RANS equations balance each other. These assumptions result in the simplified momentum and total energy equations:

\begin{equation}
    \frac{\mathrm{d}}{\mathrm{d} y_{wm}} \left[ \left( \hat{\mu}_{wm} + \mu_{t,wm} \right) \frac{\mathrm{d} \left| u_{wm} \right| }{\mathrm{d} y_{wm}} \right] = 0,
    \label{eqn:simplifiedMomentum}
\end{equation}

\begin{equation}
    \frac{\mathrm{d}}{\mathrm{d} y_{wm}} \left[ \frac{1}{\gamma-1} \left( \frac{\hat{\mu}_{wm}}{\mathrm{Pr}} + \frac{\mu_{t,wm}}{\mathrm{Pr_{t,wm}}} \right) \textcolor{black}{\frac{\mathrm{d} T_{wm}}{\mathrm{d} y_{wm}}} \right] = - \frac{\mathrm{d}}{\mathrm{d} y_{wm}} \left[ \left( \hat{\mu}_{wm} + \mu_{t,wm} \right) \left| u_{wm} \right| \frac{\mathrm{d} \left| u_{wm} \right| }{\mathrm{d} y_{wm}} \right],
    \label{eqn:simplifiedEnergy}
\end{equation}

\noindent where $\left( \cdot \right)_{wm}$ denotes that the quantity is related solely to the wall model and $\left| u_{wm} \right| = \sqrt{u^2_{wm} + w^2_{wm}}$ is the magnitude of the wall parallel velocities. \textcolor{black}{Note that throughout this section, $\left( \cdot \right)_{wm} |_{w}$ denotes wall model values at the wall.} \textcolor{black}{An eddy viscosity assumption is invoked and has the formulation:}

\begin{equation}
    \textcolor{black}{\mu_{t,wm} = \mathcal{K} \rho_{wm} \sqrt{\frac{ \left| \tau_{wm} \right|_w}{\rho_{wm}}} y_{wm} D,}
    \label{eqn:eddyViscosity}
\end{equation}

\noindent \textcolor{black}{where $D$ is the damping function. We used the van Driest damping function:}

\begin{equation}
    \textcolor{black}{D = \left[ 1 - \exp \left(-y^*/A^+ \right) \right]^2.}
\end{equation}

\noindent \textcolor{black}{Note that the wall distance in the van Driest damping function is in semi-local scaling (density and dynamic viscosity are functions of wall distance rather than constant at their wall values):}

\begin{equation}
 \textcolor{black}{y^* = \frac{y_{wm} \sqrt{\rho_{wm} \left| \tau_{wm} \right|_w }}{\hat{\mu}_{wm}}}.
\end{equation}

\noindent The parameters were taken as $\mathcal{K} = 0.41$, $A^+ = 17$, and $\mathrm{Pr}_{t,wm} = 0.9$. The dynamic viscosity was computed via Sutherland's law and was scaled in accordance with the outer LES non-dimensionalization method. Therefore, just as in the outer LES, $ \hat{\mu}_{wm} = \mu_{wm} \mathrm{Ma}/\mathrm{Re}$. Moreover, constant pressure was assumed from the wall model exchange location to the wall and the same ideal gas law was used as in the outer LES, i.e. $\rho_{wm} = \gamma p_{wm}/T_{wm}$, where $p_{wm}$ was taken to be constant.

The equations were solved on one-dimensional finite-volume grids normal to each wall cell center. Each one-dimensional grid begins at $y_{wm} = 0$ and extends up to $y_{wm} = h_{wm}$, where $h_{wm}$ is the desired height of the wall model grids. \textcolor{black}{Note that throughout this section, $\left( \cdot \right)_{wm} |_{h_{wm}}$ denotes wall model values at the wall model exchange location, whereas $\left( \cdot \right)_{LES} |_{h_{wm}}$ denotes the values from the resolved LES at the wall model exchange location.} $h_{wm}$ should be in the resolved logarithmic layer of the outer LES solution. The interface locations were obtained from: 

\begin{equation}
    \left. y_{wm} \right|_{j+\frac{1}{2}} = h_{wm} \left[ 1 - \frac{\tanh \left( \alpha_s j/N_{wm} \right)}{\tanh \left( \alpha_s \right)} \right], \quad \text{for } j \in [N_{wm}...0],
\end{equation}

\noindent where $\alpha_s$ is a stretching parameter and $N_{wm}$ is the number of cells used for the wall model grids. This stretched grid formulation has the advantage of placing the top cell interface at a height of 1 \cite{mettu2022wall}. Therefore, to ensure that the wall model grids extend up to $h_{wm}$, the cell interfaces are simply scaled accordingly. The cell centers were then computed from: 

\begin{equation}
    \left. y_{wm} \right|_{j} = \left( \left. y_{wm} \right|_{j+\frac{1}{2}} + \left. y_{wm} \right|_{j-\frac{1}{2}} \right)/2, \quad \text{for } j \in [1...N_{wm}].
\end{equation}

Eqns. \ref{eqn:simplifiedMomentum} and \ref{eqn:simplifiedEnergy} were discretized using a second-order central difference scheme in the interior, whereas at the wall and matching location, first-order forward and backward differences were used, respectively. This resulted in two tridiagonal systems of equations for which the Thomas algorithm was employed. At the wall, an isothermal viscous wall boundary condition was applied (i.e., $\left| u_{wm} \right|_{w} = 0$, $\left. T_{wm} \right|_{w} = T_{w}$), whereas at $h_{wm}$, the boundary simply matched the values of the LES at the matching location (i.e., $\left| u_{wm} \right|_{h_{wm}} = \left| u_{LES} \right|_{h_{wm}}, \left. T_{wm} \right|_{h_{wm}} = \left. T_{LES} \right|_{h_{wm}}$). Note that $h_{wm}$ does not necessarily coincide with an outer LES grid cell center. As such, inverse squared distance weighted interpolation from the 27 surrounding cells was used. 

Eqns. \ref{eqn:simplifiedMomentum} and \ref{eqn:simplifiedEnergy} were solved iteratively for the wall modelled velocity and temperature profiles at each one-dimensional finite-volume grid until the differences of the successive values of the wall modelled stresses were less than $10^{-8}$. The wall modelled wall shear stress and wall heat flux were computed from:

\begin{subequations}
    \begin{gather}
        \left. \tau_{wm} \right|_w = \left. \hat{\mu}_{wm} \right|_w \left. \frac{\mathrm{d} \left| u_{wm} \right|}{\mathrm{d} y_{wm}} \right|_{w},
        \quad
        \left. q_{wm} \right|_w = \left. \hat{\kappa}_{wm} \right|_w \left. \frac{\mathrm{d} T_{wm}}{\mathrm{d} y_{wm}} \right|_{w},
        \tag{\theequation a-\theequation b}
    \end{gather}
\end{subequations}

\noindent where,

\begin{subequations}
    \begin{gather}
        \left. \frac{\mathrm{d} \left| u_{wm} \right|}{\mathrm{d} y_{wm}} \right|_{w} = \frac{\left| u_{wm} \right|_1 - \left| u_{wm} \right|_w}{ \left. y_{wm} \right|_1 - \left. y_{wm} \right|_w},
        \quad
        \left. \frac{\mathrm{d} T_{wm}}{\mathrm{d} y_{wm}} \right|_{w} = \frac{ \left. T_{wm} \right|_1 - \left. T_{wm} \right|_{w}}{ \left. y_{wm} \right|_1 - \left. y_{wm} \right|_w}.
        \tag{\theequation a-\theequation b}
    \end{gather}
    \label{eqn:wallModelledFirstOrderWallGradients}
\end{subequations}

\noindent Once converged wall modelled stresses were obtained, we could apply the wall modelled wall gradients as Neumann conditions to the outer LES solver. First however, the wall velocity gradient was decomposed according to Bocquet et al. \cite{bocquet2012compressible}:

\begin{subequations}
    \begin{gather}
        \left. \frac{\mathrm{d} u_{wm}}{\mathrm{d} y_{wm}} \right|_{w} = \left. \frac{\mathrm{d} \left| u_{wm} \right|}{\mathrm{d} y_{wm}} \right|_{w} \frac{\left. u_{wm} \right|_{h_{wm}}}{\left| u_{wm} \right|_{h_{wm}}}, 
        \quad
        \left. \frac{\mathrm{d} w_{wm}}{\mathrm{d} y_{wm}} \right|_{w} = \left. \frac{\mathrm{d} \left| u_{wm} \right|}{\mathrm{d} y_{wm}} \right|_{w} \frac{\left. w_{wm} \right|_{h_{wm}}}{\left| u_{wm} \right|_{h_{wm}}}
        \tag{\theequation a-\theequation b}
    \end{gather}
    \label{eqn:wallModelledDecomposedVelocityGradients}
\end{subequations}

\noindent where $\left. u_{wm} \right|_{h_{wm}}$ and $\left. w_{wm} \right|_{h_{wm}}$ are the streamwise and spanwise velocities at the matching location, respectively. The velocity ratios on the right-hand-sides of Eqns. \ref{eqn:wallModelledDecomposedVelocityGradients} represent unit vectors locally aligned with their respective wall-parallel direction. Finally, the wall-normal viscous flux vectors were updated with the wall modelled velocity and temperature gradients. 

\subsection{Time Integration}

The explicit third-order total-variation-diminishing Runge-Kutta method \cite{gottlieb1998total} was used for time integration. The timestep, $\Delta t$, was computed from the CFL condition. We used both a convective and viscous analogue of the CFL condition. For all simulations, $\text{CFL} = 0.2$. The convective $\Delta t$ was computed from:

\begin{equation}
    \Delta t^c = \min \left[ \left( U + c \sqrt{\xi^{2}_{x} + \xi^{2}_{y} + \xi^{2}_{z}} \right)^{-1}, \left( V + c \sqrt{\eta^{2}_{x} + \eta^{2}_{y} + \eta^{2}_{z}} \right)^{-1}, \left( W + c \sqrt{\zeta^{2}_{x} + \zeta^{2}_{y} + \zeta^{2}_{z}} \right)^{-1} \right], 
\end{equation}

\noindent where $c = \sqrt{\gamma p/\rho}$ is the local speed of sound. The viscous $\Delta t$ was computed from:

\begin{equation}
    \Delta t^v = \frac{\hat{\nu}}{\alpha} \min \left[ \left( \xi^{2}_{x} + \xi^{2}_{y} + \xi^{2}_{z} \right)^{-1}, \left( \eta^{2}_{x} + \eta^{2}_{y} + \eta^{2}_{z} \right)^{-1}, \left( \zeta^{2}_{x} + \zeta^{2}_{y} + \zeta^{2}_{z} \right)^{-1} \right],
\end{equation}

\noindent where $\hat{\nu} = \hat{\mu}/\rho$ is the local, scaled kinematic viscosity and $\alpha = 4$ corresponds to that employed in the viscous spatial discretization method. Finally, the timestep was computed from:

\begin{equation}
    \Delta t = \text{CFL} \times \min \left( \Delta t^c, \Delta t^v \right).
\end{equation}

\subsection{Boundary Conditions}

In the current solver, a type 2 grid consistent with Laney \cite{laney1998computational} (pages 430-431) is used. As such, boundary conditions were implemented using ghost cells. Dirichlet boundary conditions were set according to:

\begin{equation} \label{eqn:dirichlet}
    \phi_{GC} = 2 \phi_{w} - \phi_{IC},
\end{equation}

\noindent where $\phi$ is an arbitary variable and $GC$, $w$, and $IC$ are ghost cell, wall value, and interior cell, respectively. Zero gradient Neumann boundary conditions were set according to:

\begin{equation} \label{eqn:zeroGradientNeumann}
    \phi_{GC} = \phi_{IC}.
\end{equation}

For WMLES cases, slightly different boundary conditions were adopted. In our solver, the ghost cells must be filled, even if wall gradients are provided by the wall model. As such, we combined Eqn. \ref{eqn:dirichlet} with a first-order finite difference approximation to fill the ghost cells when using the wall model. From a first-order finite difference approximation of the wall gradient (i.e. Eqns. \ref{eqn:wallModelledFirstOrderWallGradients}), solve for $\phi_w$. Then, use this $\phi_w$ in Eqn. \ref{eqn:dirichlet}. The result is identically the boundary condition used in the wall model, but is enforced correctly in the ghost cells using the wall modelled gradients. \textcolor{black}{For example, to compute the ghost cell temperature values, start from Eqn. \ref{eqn:wallModelledFirstOrderWallGradients}b and solve for $\left| T_{wm} \right|_w$:}

\begin{equation}
    \textcolor{black}{\left| T_{wm} \right|_w = \left| T_{wm} \right|_1 - \left. \frac{\mathrm{d} \left| T_{wm} \right|}{\mathrm{d} y_{wm}} \right|_{w} \left( \left. y_{wm} \right|_1 - \left. y_{wm} \right|_w \right).}
\end{equation}

\noindent \textcolor{black}{Substitute this into the $\phi_w$ term in Eqn. \ref{eqn:dirichlet} (and insert $T$ for $\phi$):}

\begin{equation}
    \textcolor{black}{T_{GC} = 2 \left[ \left| T_{wm} \right|_1 - \left. \frac{\mathrm{d} \left| T_{wm} \right|}{\mathrm{d} y_{wm}} \right|_{w} \left( \left. y_{wm} \right|_1 - \left. y_{wm} \right|_w \right) \right] - T_{IC}.}
\end{equation}

\noindent \textcolor{black}{In this fashion, the ghost cells are filled according to the wall modelled gradients.}

\section{Results and Discussion} \label{sec:resultsAndDiscussion}

In this section, we present results and accompanying discussions. Four schemes are compared:

\begin{enumerate}
    \item MEG-S-PRIM (described in Section 3.1.4),
    \item MEG-S-CONS (described in Section 3.1.4),
    \item MEG-C-PRIM (described in Section 3.1.5),
    \item and MEG-C-CONS (described in Section 3.1.5).
\end{enumerate}

\noindent MEG-S denotes the method presented in Chamarthi et al. \cite{chamarthi2023wave}. MEG-C denotes the method presented in this work. We also include the comparison of using primitive or conservative variable interpolation for both.

% shock entropy
% riemann
% double mach reflection
% kelvin helmholtz
% viscous shock tube re=500 and re=1000
% inviscid taylor green vortex
% flat plate

%\casesubsection{Two-Dimensional Shock Entropy Wave}
\begin{case}\label{case:shockEntropy}
    Two-Dimensional Shock Entropy Wave
\end{case}

This test case is an extension of the one-dimensional shock entropy wave test cases of Shu and Osher \cite{shu1989efficient}, and Titarev and Toro \cite{titarev2004finite}. The case features the interaction of a shockwave and high-frequency sinusodial oscillating waves, presenting a challenge for hybrid schemes. On one hand, the shock must be captured with sufficient dissipation to reduce undesirable high-frequency oscillations, while maintaining the high-frequency waves incumbent to the problem itself. The initial conditions were specified to be \cite{chamarthi2021high}:

\begin{equation}
    (\rho, u, v, p)= 
    \begin{cases}
        \left( 3.857143, 2.629369, 0, 10.3333 \right) & \text{if } x < -4, 
        \\
        \left[ 1 + 0.2 \sin \left( 10 x \cos \theta + 10 y \sin \theta \right), 0, 0, 1 \right] & \text{otherwise},
    \end{cases}
    \label{eqn:shockEntropyInitialConditions}
\end{equation}

\begin{table}[h!]
    \centering
    \caption{Boundary conditions of Case \ref{case:shockEntropy}.}
    \begin{tabular}{c c c c}
        \hline
        \hline
        $i_{min}$ & $i_{max}$ & $j_{min}$ & $j_{max}$ \\
        \hline
        Initial Conditions & Extrapolation & Zero Gradient & Zero Gradient \\
        \hline
        \hline
    \end{tabular}
    \label{tab:shockEntropyBoundaryConditions}
\end{table}

\noindent where $\theta = \pi/6$. The boundary conditions are shown in Table \ref{tab:shockEntropyBoundaryConditions}. The domain size for this case was $[x \times y] = [-5,5] \times [-1,1]$ and we used a uniform grid of $ N_x \times N_y = 400 \times 80$. The case was run until final time $t_f = 1.8$. The reference solution employed a grid of $1600 \times 320$.

\begin{figure}[h!]
    \centering
    \includegraphics[width=0.85\textwidth]{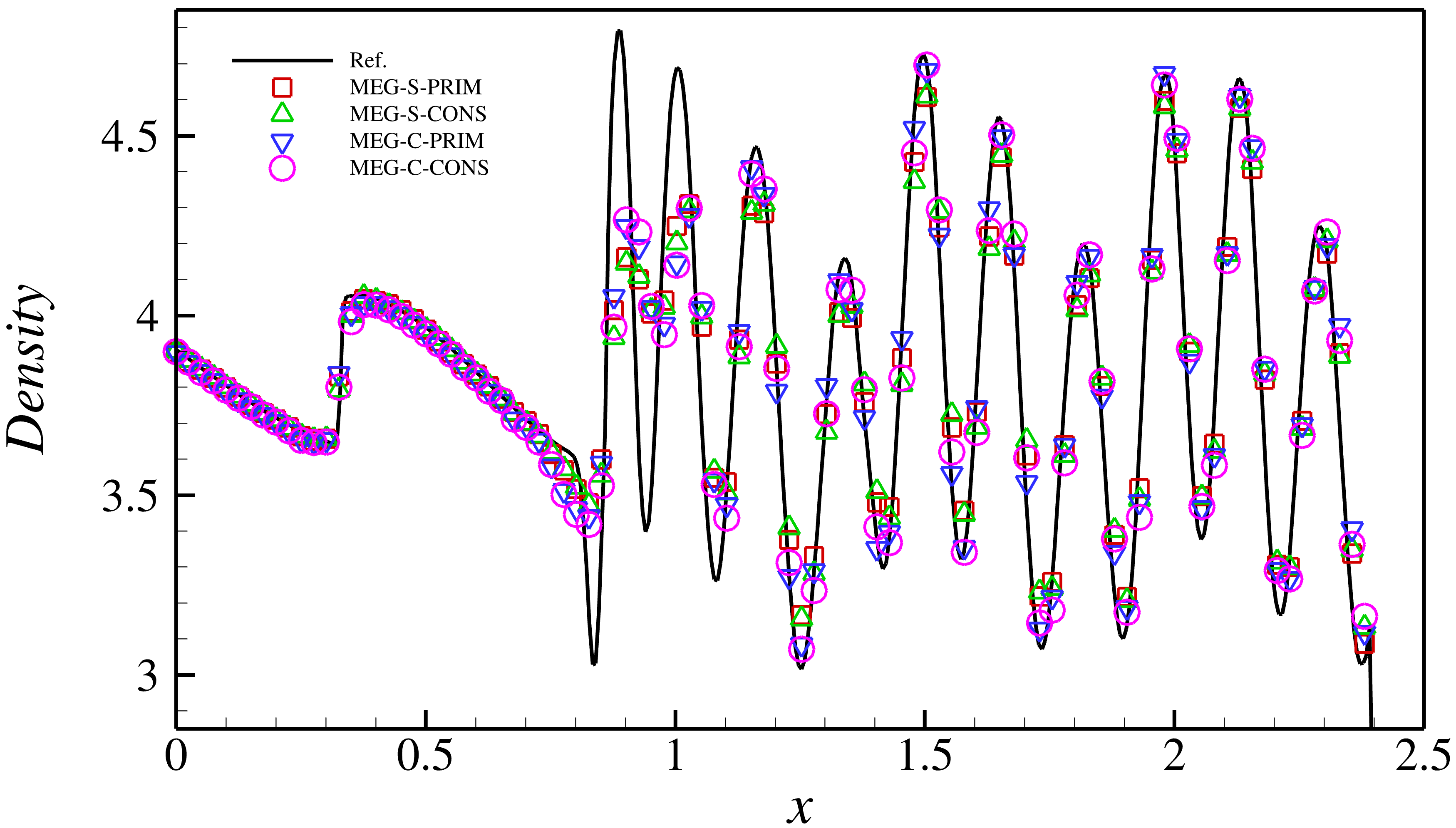}
    \caption{Density profile comparison of Case \ref{case:shockEntropy} along $y = 0$.}
    \label{fig:shockEntropy/shockEntropyDensityProfile2}
\end{figure}

Fig. \ref{fig:shockEntropy/shockEntropyDensityProfile2} shows the density profile along $y = 0$ for the considered schemes compared with the reference. It is clear that for both primitive and conservative variables, MEG-C is more capable of capturing high-frequency peaks, as a result of the reduced numerical dissipation of the method.

\begin{case}\label{case:itgv}
    Inviscid Taylor Green Vortex
\end{case}

In this test case, we evaluated the performance of the considered methods using the inviscid Taylor-Green vortex problem, in which a vortex is unsteadily decaying. The initial conditions were specified as:

\begin{equation}
    \begin{Bmatrix}
        \rho \\
        u \\
        v \\
        w \\
        p \\
    \end{Bmatrix}
    =
    \begin{Bmatrix}
        1 \\
        \sin \left( x \right) \cos \left( y \right) \cos \left( z \right) \\
        -\cos \left( x \right) \sin \left( y \right) \cos \left( z \right) \\
        0 \\
        100 + \dfrac{\left[ \cos \left( 2z \right) + 2 \right] \left[ \cos \left( 2x \right) + \cos \left( 2y \right) \right] - 2}{16}
    \end{Bmatrix}.
    \label{eqn:itgvInitialConditions}
\end{equation}

\noindent All domain boundaries were treated with periodic boundary conditions. The specific heat ratio was $\gamma = 1.4$. The domain size for this case was $[x \times y \times z] = [0,2\pi) \times [0,2\pi) \times [0,2\pi)$ and we used a uniform grid of $N_x \times N_y \times N_z = 64 \times 64 \times 64$. The case was run until final time $t_f = 10$. 

\begin{figure}[h!]
    \centering
    \begin{subfigure}{0.48\textwidth}
        \centering
        \includegraphics[width=\textwidth]{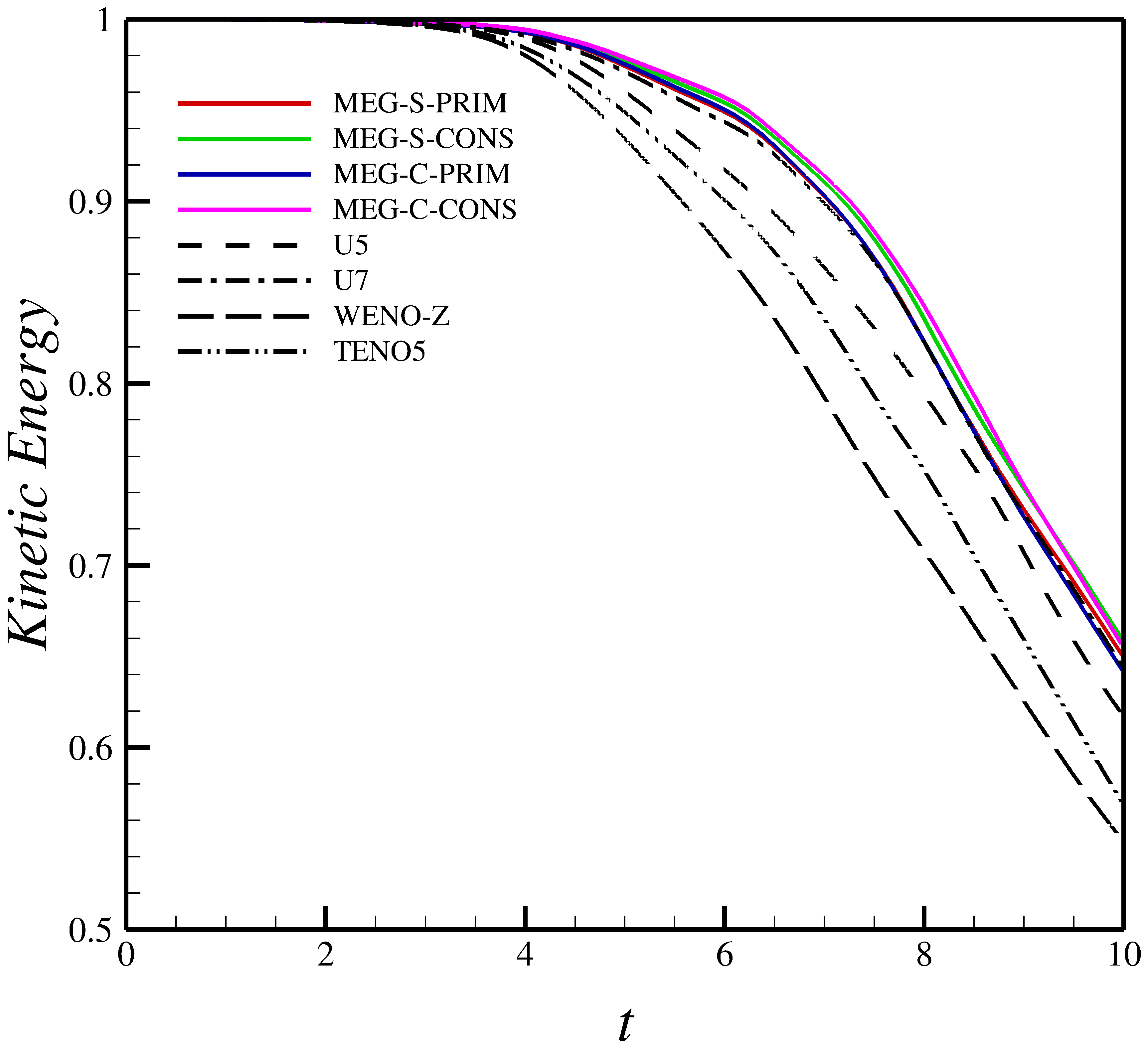}
        \caption{Normalized, volume averaged kinetic energy profiles.}
        \label{fig:itgv/afterReview_itgvKE}
    \end{subfigure}
    \hfill
    \begin{subfigure}{0.48\textwidth}
        \centering
        \includegraphics[width=\textwidth]{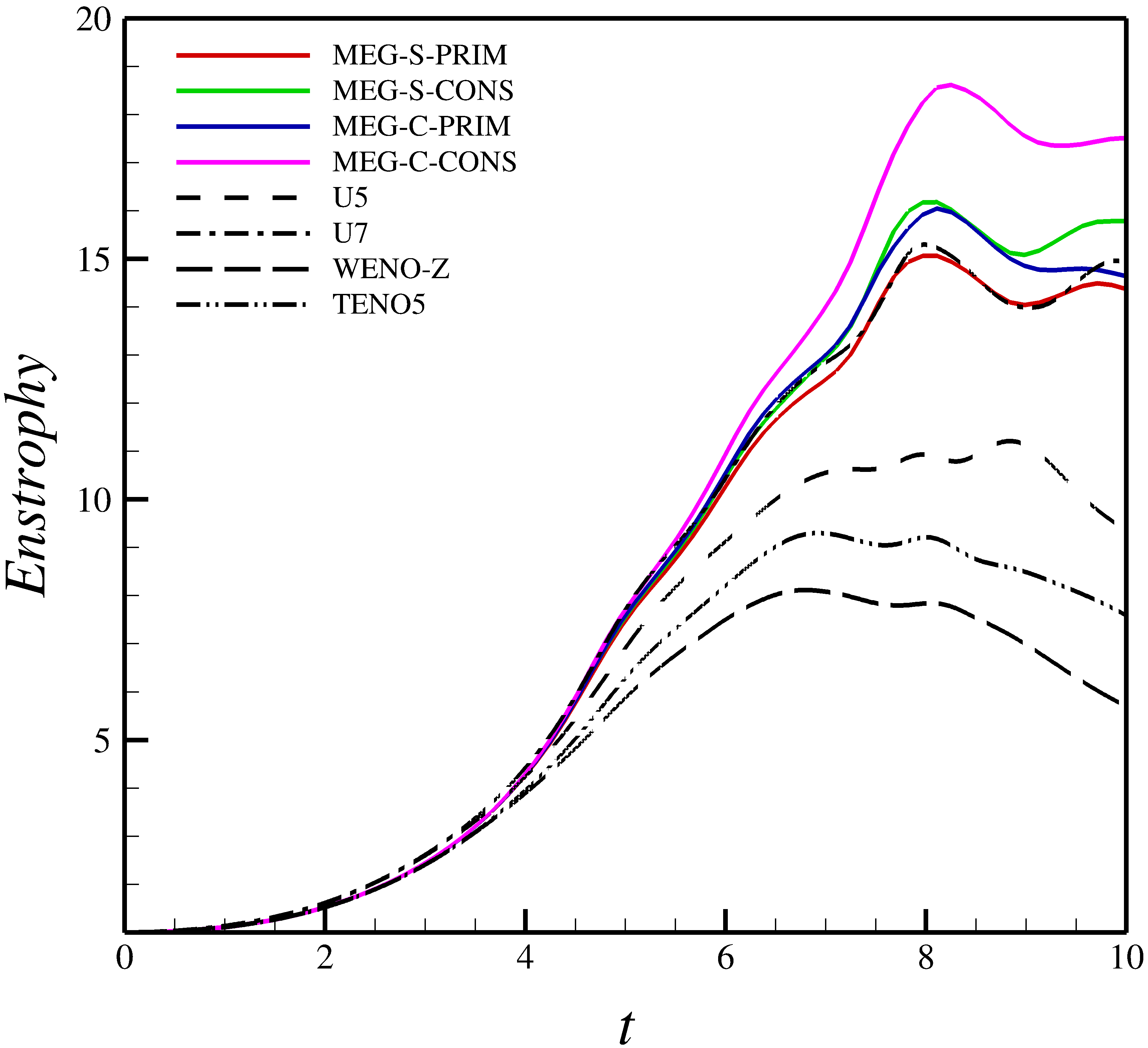}
        \caption{Normalized, volume averaged enstrophy profiles.}
        \label{fig:itgv/afterReview_itgvEnstrophy}
    \end{subfigure}
    \caption{Quantitative profiles for Case \ref{case:itgv}. Each quantity was normalized by its respective initial value.}
    \label{fig:itgvPics}
\end{figure}

Figs. \ref{fig:itgv/afterReview_itgvKE} and \ref{fig:itgv/afterReview_itgvEnstrophy} show the normalized, volume averaged kinetic energy and enstrophy profiles, respectively, of MEG-S, MEG-C\textcolor{black}{, and various other approaches}. \textcolor{black}{MEG-S and MEG-C show a similar decrease of kinetic energy with time. Notably, both MEG-S and MEG-C maintain more volume averaged kinetic energy than linear upwind fifth-order (U5) and seventh-order schemes (U7).} Observing the enstrophy profiles, it is clear that MEG-C is superior \textcolor{black}{to not only MEG-S, but other state-of-the-art hybrid numerical methods and even higher-order linear upwind schemes.} Since enstrophy is defined as half the volume integral of the squared magnitude of vorticity, we can see from the enstrophy profiles that MEG-C preserves more vortical structures than MEG-S. In fact, MEG-C performs almost as well as a fully linear implicit gradient scheme (cf. Chamarthi et al. \cite{chamarthi2023implicit}). It is clear that even though MEG-C is subject to the same limiting criteria as MEG-S, the use of centralization in characteristic space allows for reduced unnecessary numerical dissipation. Furthermore, it is important to note that the centralized approach presented in this paper outperforms the upwind scheme for this test case. Yang et al. \cite{yang2023novel} observed that an upwind TENO scheme resulted in higher enstrophy than the corresponding central TENO scheme. This is despite the fact that the central TENO scheme should provide for less kinetic energy dissipation and higher enstrophy. On the contrary, the centralized GBR method presented herein outperformed the upwind scheme, as it should, highlighting the strengths and novelty of the current approach.

%\casesubsection{Two-Dimensional Viscous Shock Tube}
\begin{case}\label{case:vst}
    Two-Dimensional Viscous Shock Tube
\end{case}

\noindent {{\textcolor{black}{\underline{$\mathrm{Re_{\infty}} = 2,500$}}}

In this test case, a shock wave and contact discontinuity propagate towards the right, causing a boundary layer to form on the bottom wall. After some time, the discontinuities impinge on the right wall, causing complex interactions with the bottom wall boundary layer as they reflect back towards the left. The initial conditions were:

\begin{equation}
    (\rho, u, v, p)= 
    \begin{cases}
        (120, 0, 0, 120/\gamma) & \text{for } 0 \leq x < 0.5, 
        \\
        (1.2, 0, 0, 1.2/\gamma) & \text{for } 0.5 \leq x < 1. 
    \end{cases}
    \label{eqn:vstInitialConditions}
\end{equation}

\noindent The relevant parameters and boundary conditions are shown in Tables \ref{tab:vstParameters} and \ref{tab:vstBoundaryConditions}, respectively. 

\begin{table}[h!]
    \centering
    \caption{Parameters of Case \ref{case:vst}.}
    \begin{tabular}{c c c c}
        \hline
        \hline
        $\mathrm{Ma}_{\infty}$ & $\mathrm{Re}_{\infty}$ & Pr & $\gamma$ \\
        \hline
        2.37 & 2,500 & 0.73 & 1.4 \\
        \hline
        \hline
    \end{tabular}
    \label{tab:vstParameters}
\end{table}

\begin{table}[h!]
    \centering
    \caption{Boundary conditions of Case \ref{case:vst}.}
    \begin{tabular}{c c c c}
        \hline
        \hline
        $i_{min}$ & $i_{max}$ & $j_{min}$ & $j_{max}$ \\
        \hline
        Adiabatic Viscous Wall & Adiabatic Viscous Wall & Adiabatic Viscous Wall & Zero Gradient \\
        \hline
        \hline
    \end{tabular}
    \label{tab:vstBoundaryConditions}
\end{table}

\noindent The domain size for this case was $[x \times y] = [0,1] \times [0,0.5]$ and we used a uniform grid of $N_x \times N_y = 2000 \times 1000$. The case was run until final time $t_f = 1$. The reference solution employed a grid totaling $109.5 \times 10^6$ cells \cite{kundu2021investigation}.

\begin{figure}[h!]
    \centering
    \includegraphics[width=0.85\textwidth]{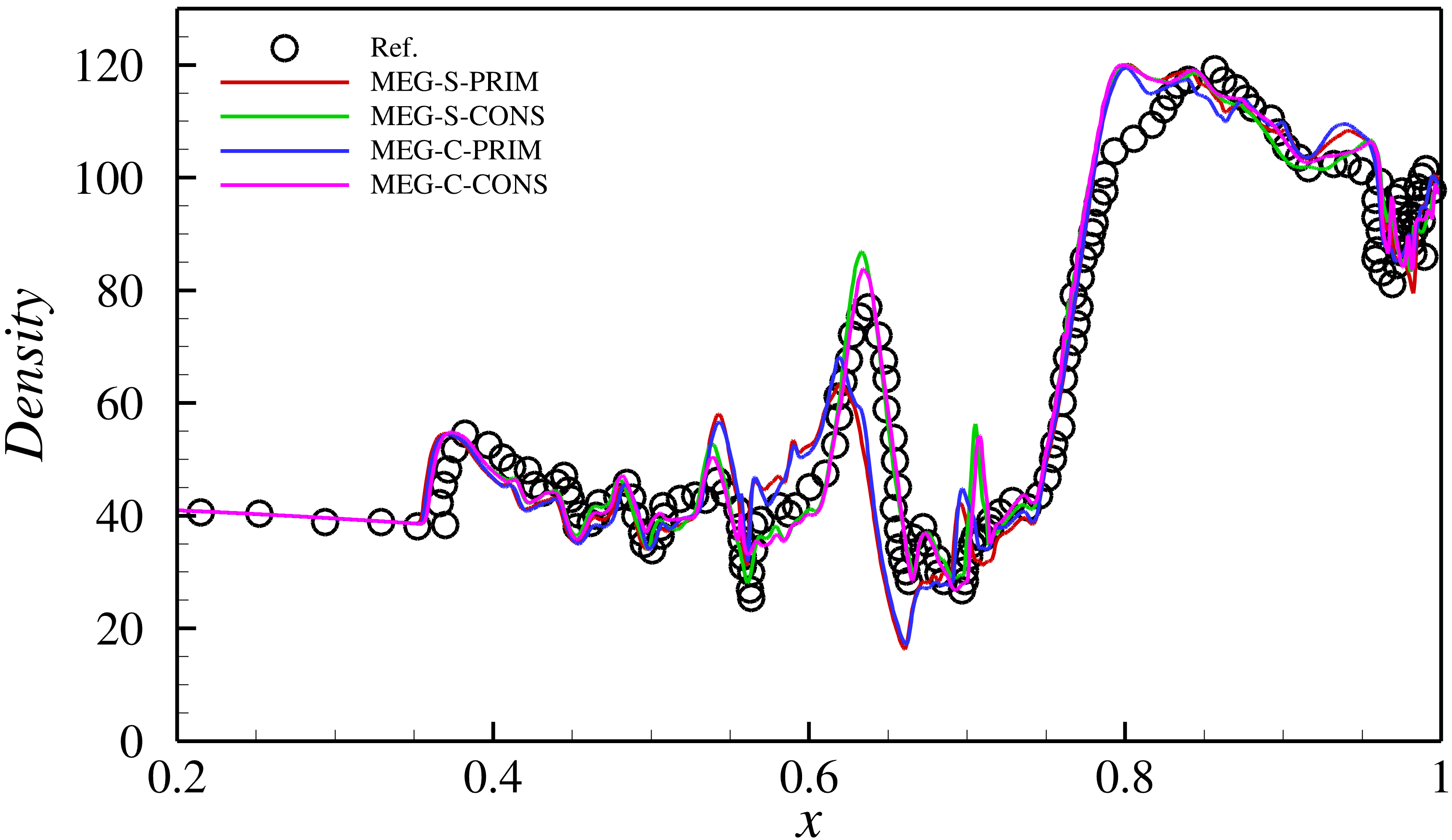}
    \caption{Wall density profile comparison of Case \ref{case:vst} at \textcolor{black}{$\mathrm{Re_{\infty}} = 2,500$}.}
    \label{fig:vst/vstWallDensityProfile2}
\end{figure}

Observing Fig. \ref{fig:vst/vstWallDensityProfile2}, while all schemes show satisfactory agreement with the reference, MEG-C-CONS shows the best overall match. \textcolor{black}{It is worth noting that at this Reynolds number, the chosen grid is under-resolved. This was done purposely to showcase the method's capabilities. However, in the following, we present grid-converged results at a lower Reynolds number.}\\

\noindent {{\textcolor{black}{\underline{$\mathrm{Re_{\infty}} = 500$}}}

\textcolor{black}{We also considered the same test case at a lower Reynolds number of $\mathrm{Re_{\infty}} = 500$, which allowed for well-converged, grid independent results at small computational cost. All simulation conditions were the same as in the $\mathrm{Re_{\infty}} = 2500$ case; however, we used a uniform grid of $N_x \times N_y = 640 \times 320$. Since this test case is only included to demonstrate the method's grid independence, we only present the results from MEG-C-CONS.} 

\begin{figure}[h!]
    \centering
    \includegraphics[width=0.85\textwidth]{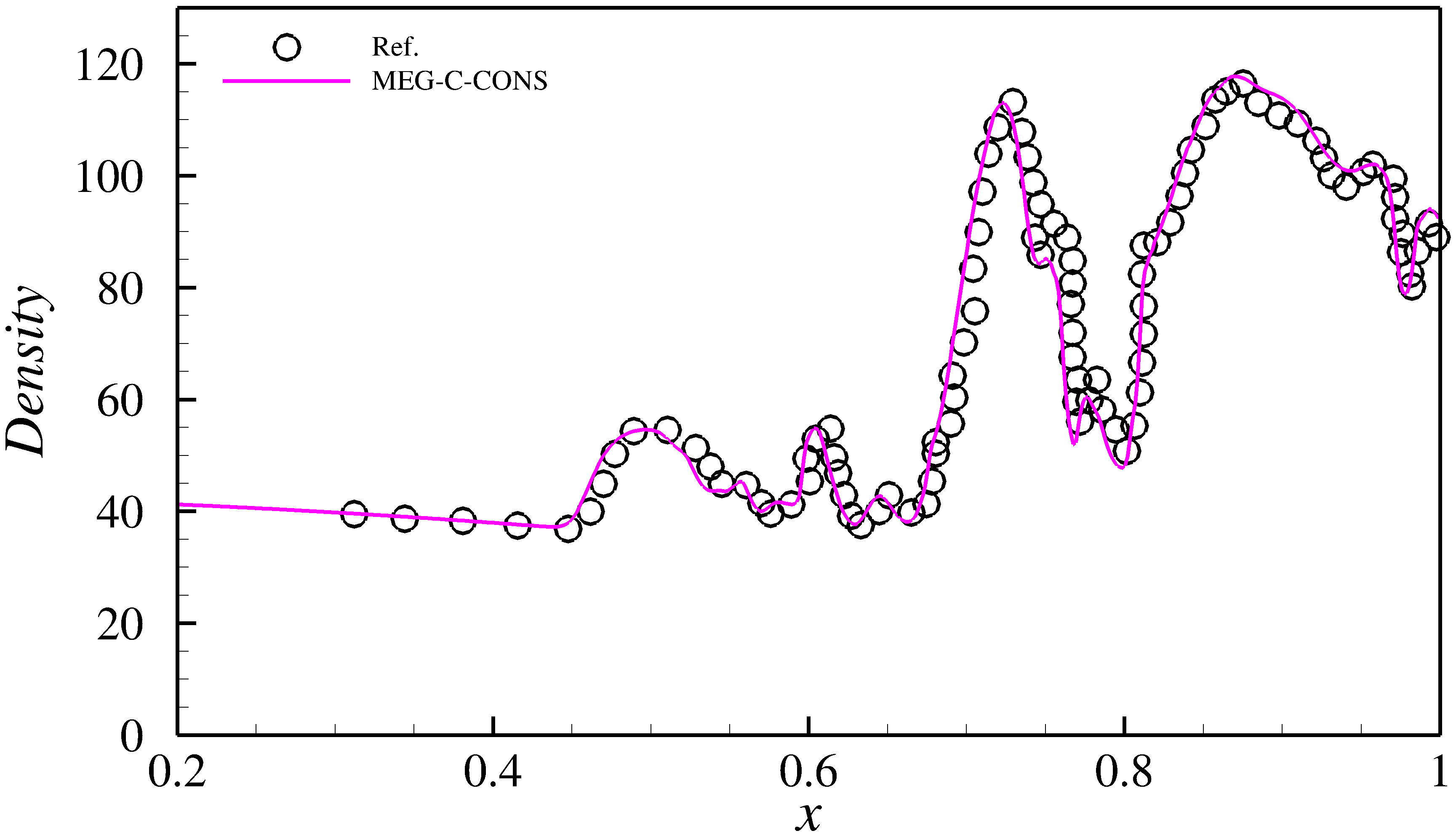}
    \caption{\textcolor{black}{Wall density profile comparison of Case \ref{case:vst} at $\mathrm{Re} = 500$.}}
    \label{fig:vst/afterReview_vstRe500WallDensityProfile}
\end{figure}

\noindent \textcolor{black}{The reference grid size for this test case was $N_x \times N_y = 1280 \times 640$ \cite{chamarthi2022importance}. Observing Fig. \ref{fig:vst/afterReview_vstRe500WallDensityProfile}, MEG-C-CONS matches the reference excellently.}

%\casesubsection{Oblique Shock Impingement from $4^{\circ}$ Wedge on Mach 6 Disturbed Laminar Boundary Layer}
\begin{case}\label{case:sbli}
    Oblique Shock Impingement from $4^{\circ}$ Wedge on Mach 6 Disturbed Laminar Boundary Layer
\end{case}

\begin{table}[h!]
    \centering
    \caption{Non-dimensional parameters of Case \ref{case:sbli}.}
    \begin{tabular}{c c c c c}
        \hline
        \hline
        $\mathrm{Ma}_{\infty}$ & $\mathrm{Re}_{\infty,\delta^{*}_{0}}$ & $\mathrm{Re}_{\infty,x_{0}}$ & Pr & $\gamma$ \\[3pt]
        \hline
        6 & 6,830 & 314,252 & 0.72 & 1.4 \\
        \hline
        \hline
    \end{tabular}
    \label{tab:sbliNondimensionalParameters}
\end{table}

\begin{table}[h!]
    \centering
    \caption{Reference values of Case \ref{case:sbli}.}
    \begin{tabular}{c c c c c}
        \hline
        \hline
        $T_{\infty}$, \SI{}{\kelvin} & $T_{w}$, \SI{}{\kelvin} & $\rho_{\infty}$, \SI{}{\kilogram\per\meter\cubed} & $\mu_{\infty}$, \SI{}{\pascal\second} & $u_{\infty}$, \SI{}{\meter\per\second} \\
        \hline
        65 & 292.5 & 0.0267 & $4.16 \times 10^{-6}$ & 969.69 \\
        \hline
        \hline
    \end{tabular}
    \label{tab:sbliReferenceValues}
\end{table}

\begin{table}[h!]
    \centering
    \caption{Boundary conditions of Case \ref{case:sbli}. RH refers to Rankine-Hugoniot jump conditions.}
    \begin{tabular}{c c c c c}
        \hline
        \hline
        $i_{min}$ & $i_{max}$ & $j_{min}$ & $j_{max}$ & $k_{min}$, $k_{max}$ \\
        \hline
        Freestream & Extrapolation & Isothermal Viscous Wall & RH Conditions &  Periodic \\
        \hline
        \hline
    \end{tabular}
    \label{tab:sbliBoundaryConditions}
\end{table}

\begin{table}[h!]
    \centering
    \caption{Grid details of Case \ref{case:sbli}.}
    \begin{tabular}{c c c c}
        \hline
        \hline
        Domain Size & Grid & Total Grid Size & $ \left.\Delta y \right|_w$\\
        \hline
        $300 \delta^{*}_{0} \times 25 \delta^{*}_{0} \times 45 \delta^{*}_{0}$ & $588 \times 64 \times 110$ & 4M & $0.204 \delta^{*}_{0}$ \\
        \hline
        \hline
    \end{tabular}
    \label{tab:sbliGridDetails}
\end{table}

The next case considered was oblique shock impingement on a Mach 6 disturbed laminar boundary layer. This case was studied experimentally \cite{schulein2014effects}, \cite{willems2015experiments}, experimentally and with DNS \cite{sandham2014transitional}, as well as with WMLES \cite{yang2018aerodynamic}, \cite{mettu2018wall}, \cite{ganju2021progress}. In this flow system, an oblique shock caused by a $4^{\circ}$ wedge atop the domain is made to impinge on a disturbed laminar boundary layer. The shock impingement causes a separation bubble to form, introducing shear layer instabilities to the flow, which interact with existing second mode disturbances propagating from the upstream flow. These instabilities eventually destabilize post-reattachment, causing boundary layer transition to turbulence. The use of a grid appropriate for WMLES implicitly applies a spatial filter to the flow due to the intentionally unresolved nature of the simulation. As such, any additional numerical or unphysical dissipation can completely change the characterization of the flow and thus must be minimized. Therefore, this test case presents a difficult and relevant system for the considered numerical schemes.

The non-dimensional parameters for this case may be found in Table \ref{tab:sbliNondimensionalParameters}. Note that the Mach number is based on the freestream velocity and temperature. The Reynolds number is based on the displacement thickness at a distance from the flat plate leading edge, $Re_{x_{0}} = 314,252$ or $x = 46 \delta^{*}_{0}$. The reference values and boundary conditions for this case can be found in Tables \ref{tab:sbliReferenceValues} and \ref{tab:sbliBoundaryConditions}, respectively. The initial condition of this case was set according to a compressible similarity solution at $x = 46 \delta^{*}_{0}$ downstream of the flat plate leading edge. This initial condition was computed from the MATLAB code provided by Oz and Kara \cite{oz2021cfd}.

The oblique shock impingement on the boundary layer is insufficient to cause boundary layer transition alone. This was noted in the original DNS study of Sandham et al. \cite{sandham2014transitional}. However, in experiments of the same case, boundary layer transition was observed. This is believed to be a result of wind tunnel freestream disturbances in addition to instabilities caused by the shock boundary layer interaction. As such, in the DNS study, freestream disturbances were added to the density at $x = 0 \delta^{*}_{0}$ in an attempt to characterize the freestream disturbances present in the experiments. These disturbances are of the form:

\begin{equation}
    \rho' = A W(y) \sum^{J}_{j=1} \cos \left( 2\pi j z/L_z + \phi_j \right) \sum^{K}_{k=1} \sin \left( 2\pi f_k t + \psi_k \right), 
\end{equation}

\noindent where $W(y) = 1 - \exp \left( -y^3 \right)$ is a window function to dampen the disturbances in the boundary layer, $J = 16$ and $K = 20$ are cutoff wavenumbers, $f_k = 0.02 k$ is the frequency, and $\phi_j$ and $\psi_k$ are random phases in $\left[ 0,2 \pi \right]$. The disturbance amplitude, $A$, is tuned based on the grid resolution considered. For example, in the DNS of Sandham et al. \cite{sandham2014transitional}, using approximately 200 million cells, $A = 0.0005$. Whereas in the subsequent WMLES studies of Yang et al. \cite{yang2018aerodynamic}, Mettu and Subbareddy \cite{mettu2018wall}, and Ganju et al. \cite{ganju2021progress}, using between one and four million cells, $A = 0.001$. Thus for this study, we took $A = 0.001$. The grid information may be found in Table \ref{tab:sbliGridDetails}. Note that the grid was made according to WMLES standards conveniently defined at the WMLES Resource \cite{wmles2023}. The wall model exchange location was placed at $h_{wm} = 0.1 \delta$, where $\delta = 5.1 \delta^{*}_{0}$. Moreover, we used $N_{wm} = 100$, with $\alpha_s = 3.0$. This placed $\left. y^{+}_{wm} \right|_{w} < 1$. \textcolor{black}{Note that $h_{wm}$ was chosen based on the WMLES standards defined at the WMLES Resource \cite{wmles2023}. In addition, this $h_{wm}$ was used in the previous WMLES studies of this case by Yang et al. \cite{yang2018aerodynamic}, Mettu and Subbareddy \cite{mettu2018wall}, and Ganju et al. \cite{ganju2021progress}}.

\begin{figure}[h!]
    \centering
    \includegraphics[width=\textwidth]{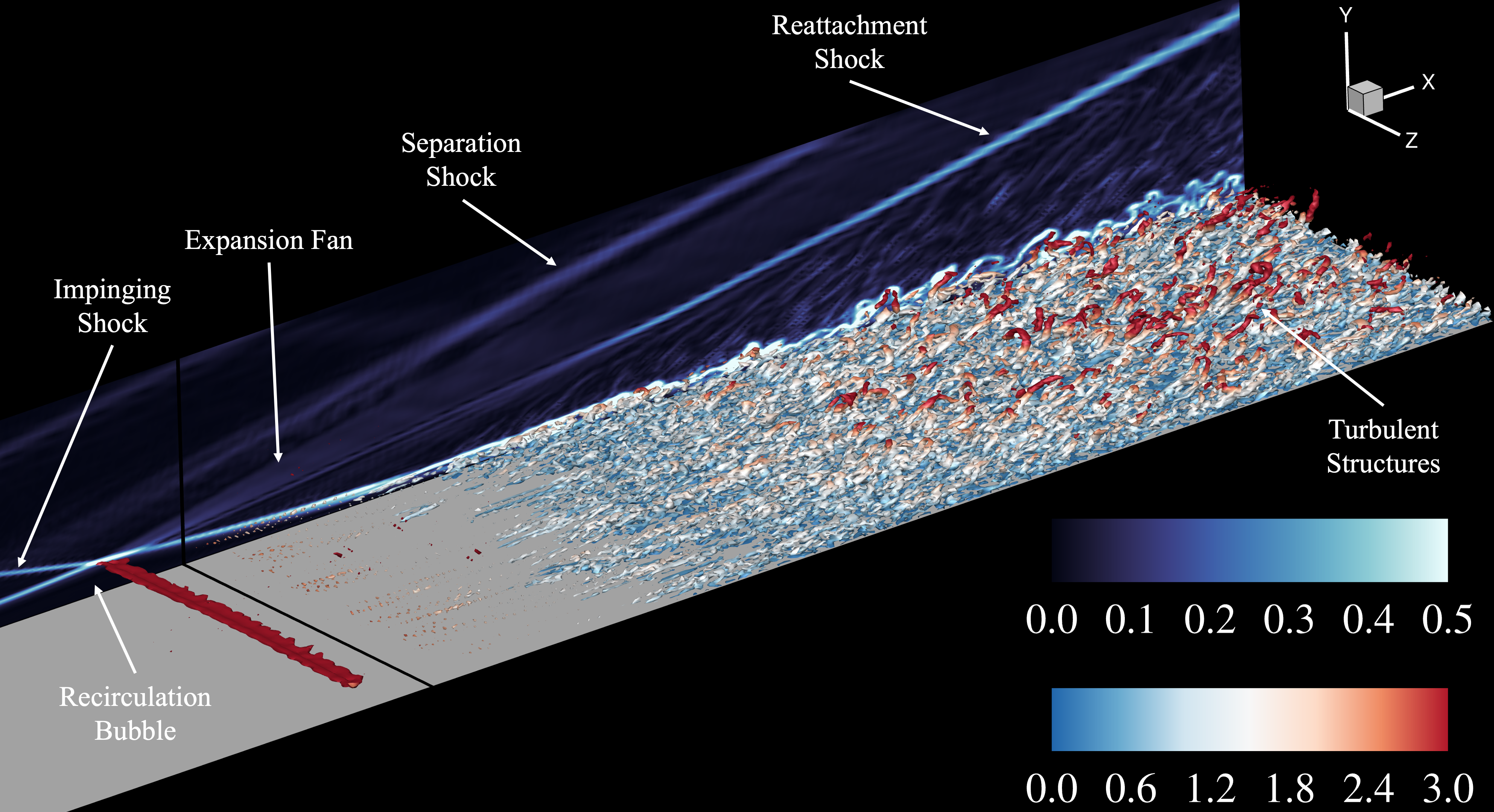}
    \caption{Qualitative figure of Case \ref{case:sbli}. $x$-$y$ slice: instantaneous density gradient magnitude contour. Iso-surfaces: Q-criterion ($Q = 0.1$) colored by wall-normal distance.}
    \label{fig:sbli/sbliQualitativePicAnnotated}
\end{figure}

\begin{figure}[h!]
    \centering
    \begin{subfigure}[t]{\textwidth}
        \centering
        \includegraphics[width=\textwidth]{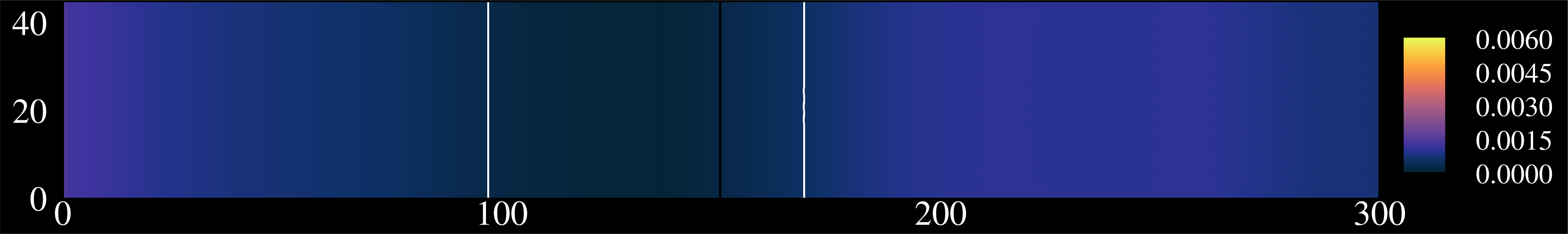}
        \caption{MEG-S-PRIM.}
        \label{fig:sbli/sbliStContourMEG-S-PRIM}
    \end{subfigure}
    \vfill
    \begin{subfigure}[t]{\textwidth}
        \centering
        \includegraphics[width=\textwidth]{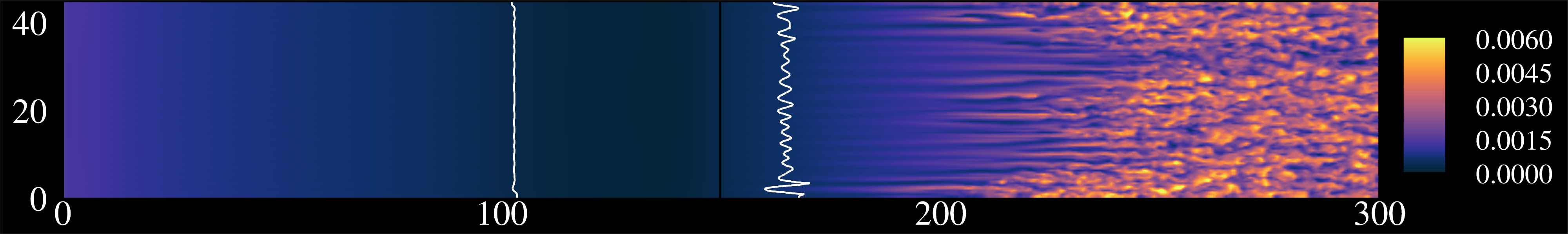}
        \caption{MEG-S-CONS.}
        \label{fig:sbli/sbliStContourMEG-S-CONS}
    \end{subfigure}
    \vfill
    \begin{subfigure}[t]{\textwidth}
        \centering
        \includegraphics[width=\textwidth]{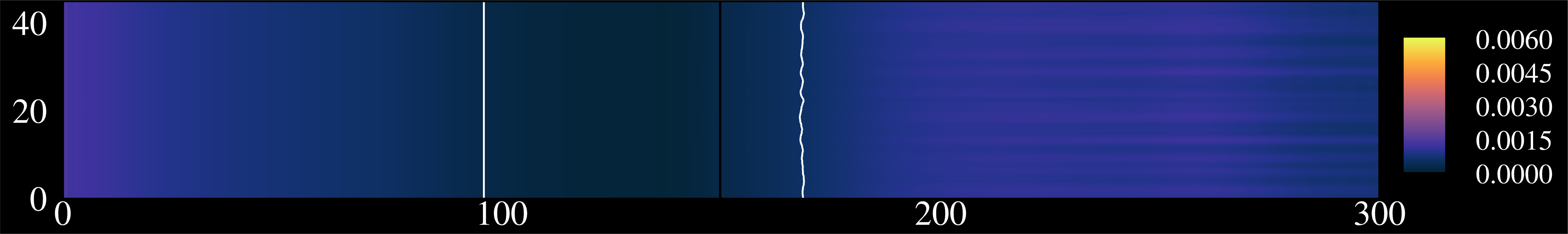}
        \caption{MEG-C-PRIM.}
        \label{fig:sbli/sbliStContourMEG-C-PRIM}
    \end{subfigure}
    \vfill
    \begin{subfigure}[t]{\textwidth}
        \centering
        \includegraphics[width=\textwidth]{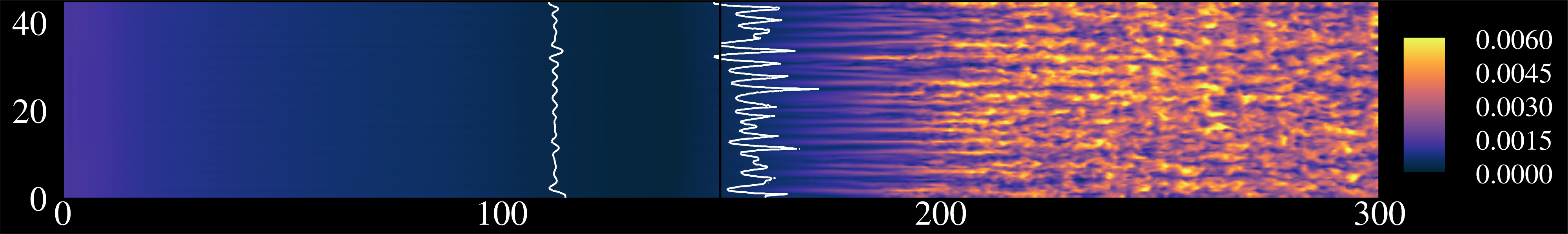}
        \caption{MEG-C-CONS.}
        \label{fig:sbli/sbliStContourMEG-C-CONS}
    \end{subfigure}
    \caption{Instantaneous Stanton number contours for the considered schemes. White lines represent surfaces of zero skin friction coefficient ($\mathrm{C}_f = 0$).}
    \label{fig:sbliStContour}
\end{figure}

Fig. \ref{fig:sbli/sbliQualitativePicAnnotated} shows a qualitative view of the instantaneous flow system using the MEG-C-CONS scheme. The shock impingement causes an adverse pressure gradient, which forms a recirculation bubble. From this bubble, a separation shock forms along with an expansion fan at the top of the bubble. Boundary layer reattachment is associated with a strong reattachment shock along with streamwise streaks. The streamwise streaks cause intense spanwise periodic wall heating. These streaks eventually destabilize and boundary layer transition to turbulence occurs shortly after. Note that this was not the scenario for all schemes. In fact, this only occurred completely for the MEG-C-CONS scheme. 

Transition to turbulence is accompanied by intense wall heating in cold-wall isothermal flows. A quantitative measure for this is the Stanton number, computed via:

\begin{equation}
    \mathrm{St} = \frac{ \left. q_{wm} \right|_w}{\rho_{\infty} u_{\infty} \mathrm{c_p} \left( T_r - T_w \right)},
\end{equation}

\noindent where $T_r = \left[ 1 + r \left( \gamma - 1 \right) \mathrm{Ma}^2_{\infty}/2 \right]$ is the recovery temperature, $r = \mathrm{Pr}^{1/2} = 0.83$ is the recovery factor, and $\rho_{\infty} = 1$, $u_{\infty} = \mathrm{Ma}$, and $\mathrm{c_p} = \left( \gamma - 1 \right)^{-1}$ due to non-dimensionalization. Figs. \ref{fig:sbliStContour} shows instantaneous Stanton number contours for all considered schemes at the same time. It is clear that transition to turbulence only occurs for MEG-S-CONS and MEG-C-CONS. However, the use of conservative variable interpolation is insufficient; a qualitatively (and quantitatively) accurate flow was only achieved by MEG-C-CONS, portraying the importance of eliminating unphysical numerical dissipation due to the upwinding performed in the MEG-S scheme.

\begin{figure}[h!]
    \centering
    \includegraphics[width=0.54\textwidth]{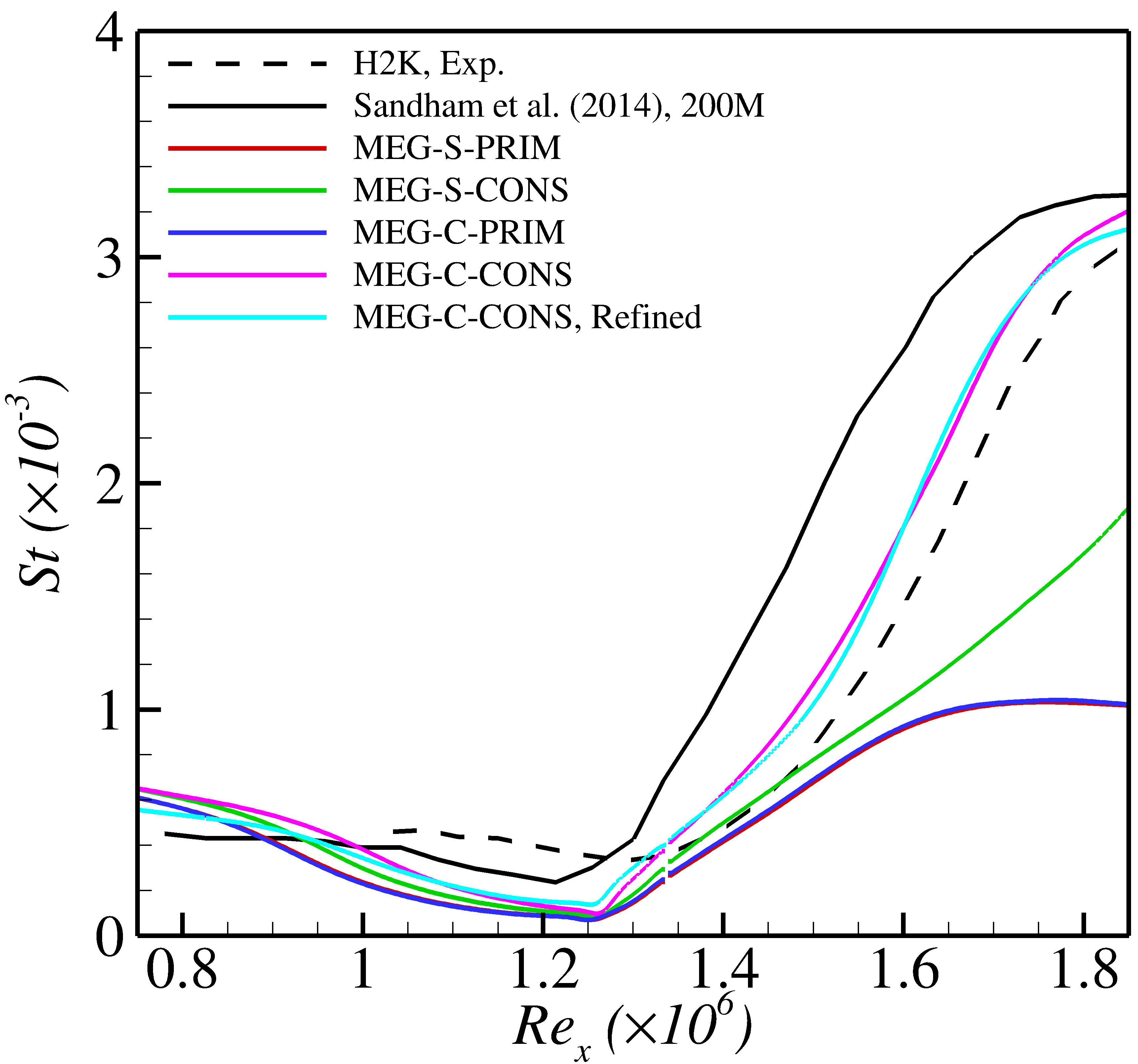}
    \caption{Time and spanwise averaged Stanton number profiles compared with \textcolor{black}{the H2K experiment \cite{willems2015experiments} and DNS results of Sandham et al. \cite{sandham2014transitional}.}}
    \label{fig:sbli/afterReview_sbliStProfile}
\end{figure}

Fig. \ref{fig:sbli/afterReview_sbliStProfile} displays the time and spanwise averaged Stanton number as a function of $\mathrm{Re}_x = x \mathrm{Re}_{\infty, \delta^{*}_{0}} + \mathrm{Re}_{x_{0}}$. A quantitative comparison was made to \textcolor{black}{the H2K experiment \cite{willems2015experiments} and DNS results of Sandham et al. \cite{sandham2014transitional}. Note that the slight mismatch in the Stanton number dip for the H2K profile was noted in the reference work of Sandham et al. \cite{sandham2014transitional}. We also include the profile from a finer-grid simulation using MEG-C-CONS. For the finer-grid case, we doubled the amount of cells in each direction (see Table \ref{tab:sbliGridDetails}), which totaled to a grid size of 33M cells.} Consistent with the qualitative portrayal in Figs. \ref{fig:sbliStContour}, the Stanton number profiles for MEG-S-PRIM and MEG-C-PRIM show only a very small increase in wall heating post-reattachment because there was no transition to turbulence. MEG-S-CONS does show a large increase in Stanton number post-impingement, however the profile shows that the wall heating peak was not reached, indicating that the flow did not fully transition to turbulence along the considered streamwise distance. MEG-C-CONS, however shows a much sharper increase in Stanton number post-reattachment, as well as good agreement with \textcolor{black}{experiment and DNS}.

\begin{remark}\label{remark:which-variable}
    In Chamarthi \cite{chamarthi2023efficient}, simulations were conducted using both primitive and conservative variables with the GBR approach. The choice of variables and test case led to notable differences in solution quality, as assessed by various metrics. Similar differences were evident when using other well-known schemes (WENO, MP5, and so forth). While the reasons behind these discrepancies are likely substantial, they are not the focus of this paper.
\end{remark}

%\casesubsection{Hypersonic Transition to Turbulence over a $15^{\circ}$ Compression Ramp}
\begin{case}\label{case:ramp}
    Mach 7.7 Transition to Turbulence over a $15^{\circ}$ Compression Ramp
\end{case}

\begin{figure}[h!]
    \centering
    \includegraphics[width=\textwidth]{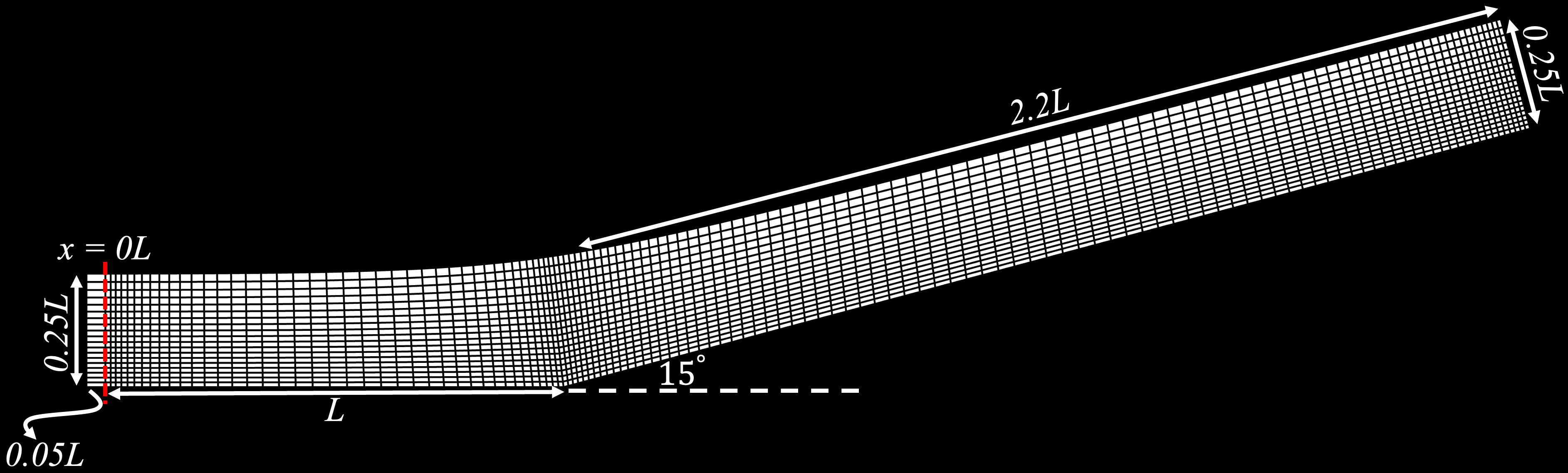}
    \caption{Grid and relevant dimensions of Case \ref{case:ramp}. Every tenth point is shown for clarity. The flat plate leading edge is at $x = 0L$. The grid is lightly clustered in the wall-normal direction and in the streamwise direction near the leading edge, corner, and outlet. Care was taken to ensure that near wall cells were orthogonal, however this is not depicted here since every tenth point is shown.}
    \label{fig:ramp/rampSchematic}
\end{figure}

\begin{table}[h!]
    \centering
    \caption{Non-dimensional parameters of Case \ref{case:ramp}.}
    \begin{tabular}{c c c c c}
        \hline
        \hline
        $\mathrm{Ma}_{\infty}$ & $\mathrm{Re}_{\infty,L}$ & Pr & $\gamma$ \\
        \hline
        7.7 & $8.6 \times 10^5$ & 0.71 & 1.4 \\
        \hline
        \hline
    \end{tabular}
    \label{tab:rampNondimensionalParameters}
\end{table}

\begin{table}[h!]
    \centering
    \caption{Reference values of Case \ref{case:ramp}.}
    \begin{tabular}{c c c c c}
        \hline
        \hline
        $T_{\infty}$, \SI{}{\kelvin} & $T_{w}$, \SI{}{\kelvin} & $\rho_{\infty}$, \SI{}{\kilogram\per\meter\cubed} & $\mu_{\infty}$, \SI{}{\pascal\second} & $u_{\infty}$, \SI{}{\meter\per\second} \\
        \hline
        125 & 293 & 0.0432 & $8.7 \times 10^{-6}$ & 1726 \\
        \hline
        \hline
    \end{tabular}
    \label{tab:rampReferenceValues}
\end{table}

\begin{table}[h!]
    \centering
    \caption{Boundary conditions of Case \ref{case:ramp}.}
    \begin{tabular}{c c c c c c}
        \hline
        \hline
        $i_{min}$ & $i_{max}$ & $j_{min}$, $x \leq 0 L$ & $j_{min}$, $x > 0 L$ & $j_{max}$ & $k_{min}$, $k_{max}$ \\
        \hline
        Freestream & Extrapolation & Inviscid Wall & Isothermal Viscous Wall & Freestream &  Periodic \\
        \hline
        \hline
    \end{tabular}
    \label{tab:rampBoundaryConditions}
\end{table}

\begin{table}[h!]
    \centering
    \caption{Grid details of Case \ref{case:ramp}.}
    \begin{tabular}{c c c}
        \hline
        \hline
        Grid & Total Grid Size & $ \left.\Delta y \right|_w$\\
        \hline
        $1135 \times 224 \times 134$ & 33.6M & $0.001 L$ \\
        \hline
        \hline
    \end{tabular}
    \label{tab:rampGridDetails}
\end{table}

The final test case considered was hypersonic transition to turbulence over a $15^{\circ}$ compression ramp. This configuration was investigated using DNS in Cao et al. \cite{cao2022transition}. The DNS study was part of an investigation examining the experimental configuration of Roghelia et al. \cite{roghelia2017experimentala,roghelia2017experimentalb}. Cao et al. \cite{cao2021unsteady} performed DNS for the exact experimental configuration but extended the case to allow for full transition to turbulence by increasing the Reynolds number and the length of the ramp \cite{cao2022transition}. This case is presented to display the use of MEG-C-CONS in a more complex scenario; therefore, the other schemes are not considered here. The non-dimensional parameters, reference values, boundary conditions, and grid details for this case may be found in Tables \ref{tab:rampNondimensionalParameters}, \ref{tab:rampReferenceValues}, \ref{tab:rampBoundaryConditions}, and \ref{tab:rampGridDetails}, respectively. The Reynolds number was based on the flat plate length, $L = \SI{0.1}{\meter}$. The grid was designed in accordance with WMLES standards so that grid spacings were based on the boundary layer thickness at the outlet, $\delta_{outlet} = 0.028L$, such that $\left( \Delta x \right.$, $\Delta y$, $\left. \Delta z \right) = \left( 0.1 \delta_{outlet} \right.$, $0.04 \delta_{outlet}$, $\left. 0.08 \delta_{outlet} \right)$. A two-dimensional simulation was run to bring about a quasi-steady mean flow that was used as an initial condition for a subsequent three-dimensional simulation. A grid of $1135 \times 224$ was used for the two-dimensional simulation. For the three-dimensonal simulation, the spanwise length was $L_z = 0.3 L$.

\begin{figure}[h!]
    \centering
    \includegraphics[width=0.8\textwidth]{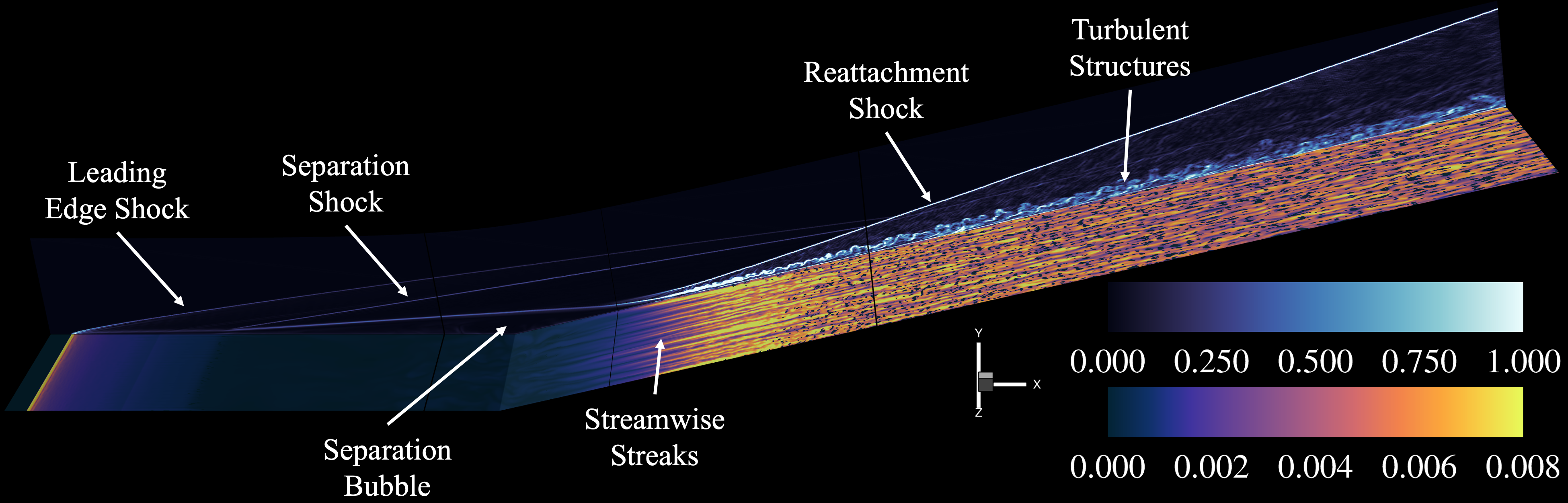}
    \caption{Qualitative figure of Case \ref{case:ramp}. $x$-$y$ slice: instantaneous density gradient magnitude contour. $x$-$z$ slice: instantaneous Stanton number contour.}
    \label{fig:ramp/rampQualitativePic3Annotated}
\end{figure}

\begin{figure}[h!]
    \centering
    \includegraphics[width=0.8\textwidth]{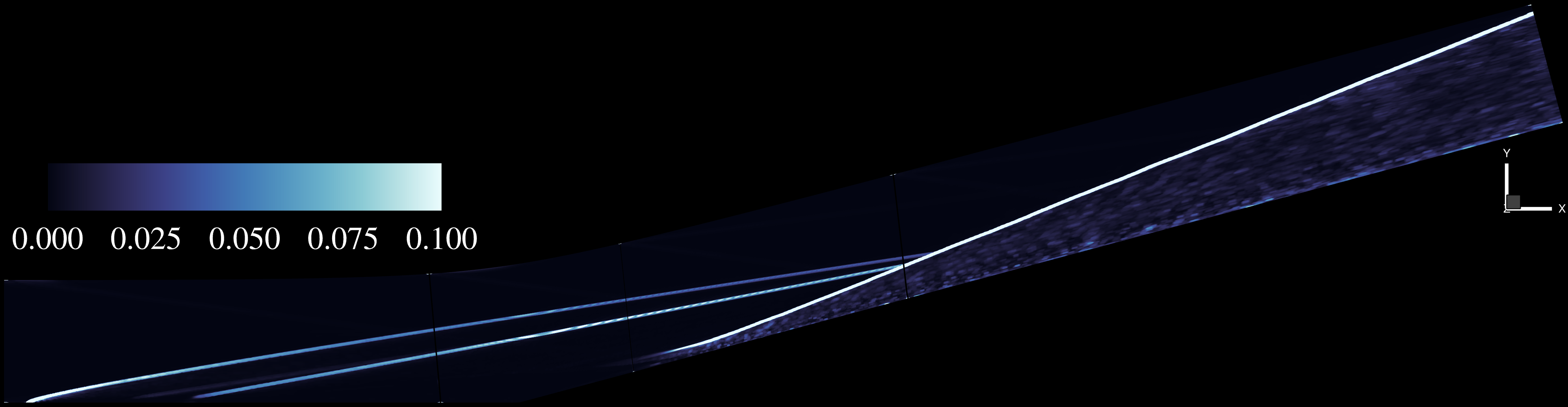}
    \caption{Instantaneous contour of $\overline{\Omega^d_{i,j,k}}$.}
    \label{fig:ramp/rampBlankedSensorPic}
\end{figure}

\begin{figure}[h!]
    \centering
    \includegraphics[width=0.8\textwidth]{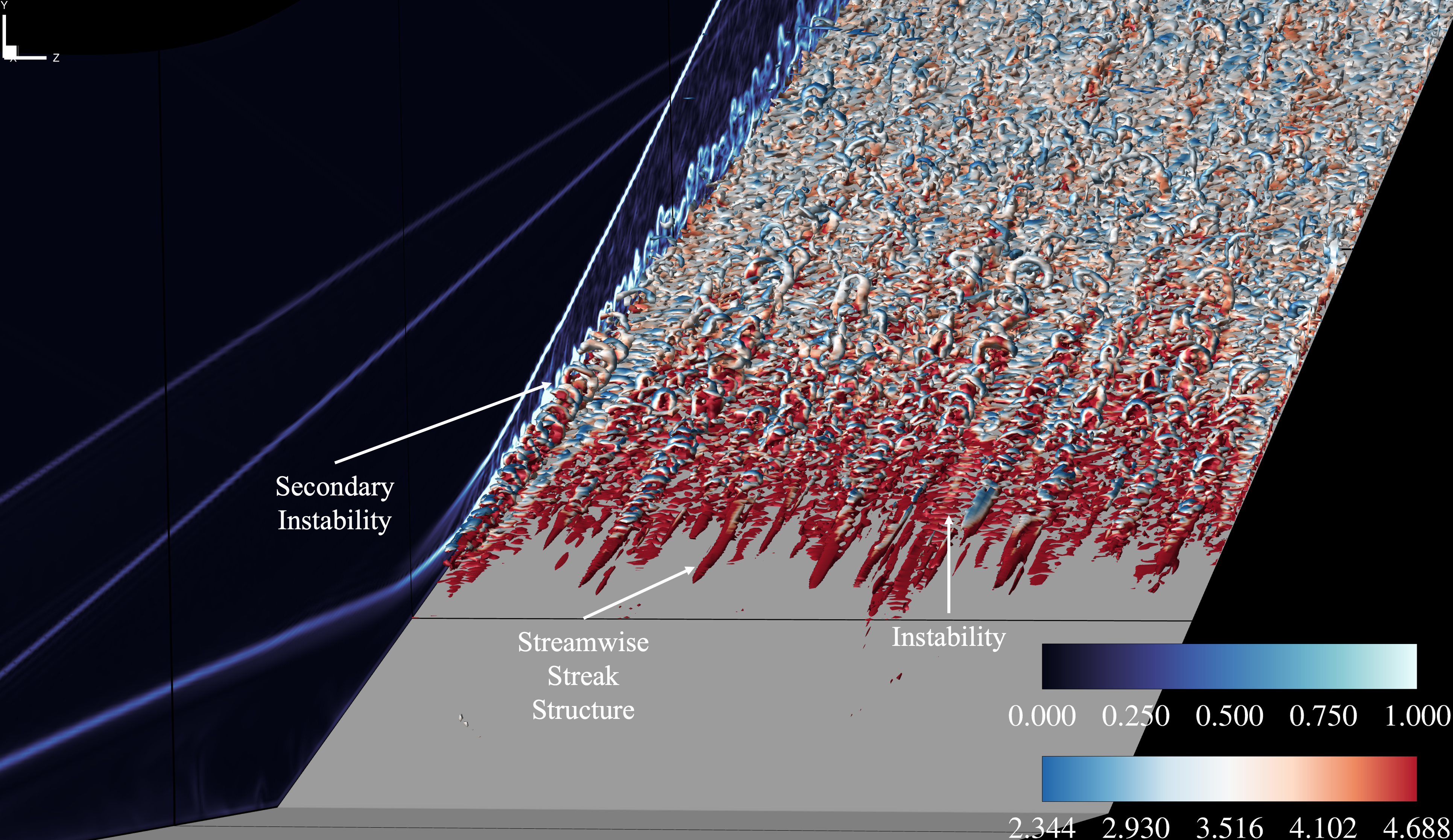}
    \caption{Zoomed in qualitative figure of Case \ref{case:ramp}. $x$-$y$ slice: instantaneous density gradient magnitude contour. Iso-surfaces: Q-criterion ($Q = 0.1$) colored by temperature.}
    \label{fig:ramp/rampQualitativePicCloseUp3Annotated}
\end{figure}

\begin{figure}[h!]
    \centering
    \includegraphics[width=0.6\textwidth]{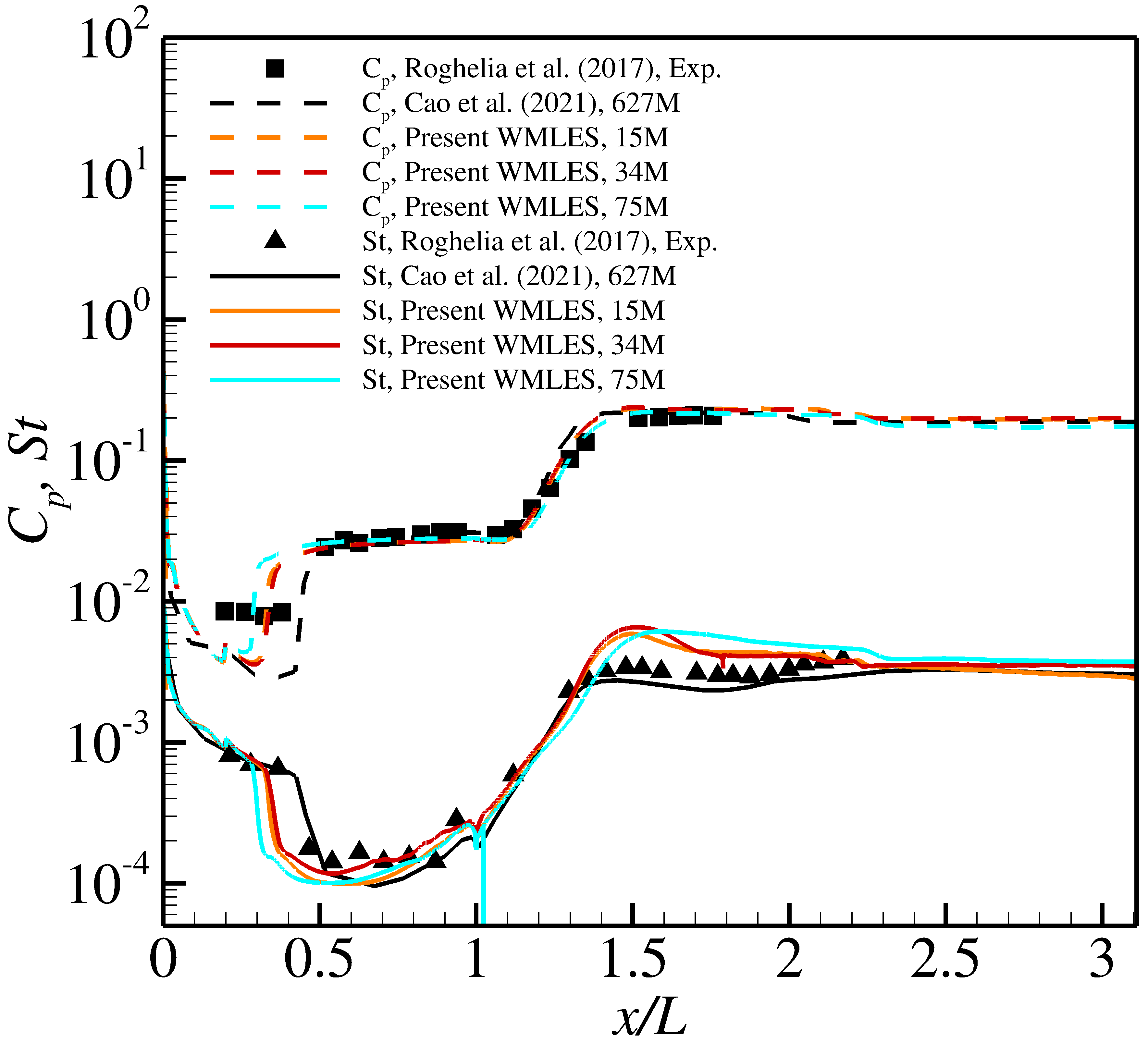}
    \caption{Time and spanwise averaged Stanton number and wall pressure coefficient compared with data from Cao et al. \cite{cao2022transition} and Roghelia et al. \cite{roghelia2017experimentala}.}
    \label{fig:ramp/afterReview_rampCpStProfile}
\end{figure}

Fig. \ref{fig:ramp/rampQualitativePic3Annotated} shows the instantaneous flow field. Slightly downstream of the inlet, a leading edge shock forms where the no-slip condition starts. Downstream, the flow is compressed from the ramp, causing a very large adverse pressure gradient, resulting in a large separation bubble and \textcolor{black}{its} associated shock system. Fig. \ref{fig:ramp/rampBlankedSensorPic} shows the instantaneous contour of $\overline{\Omega^d_{i,j,k}}$ displaying the capability of the sensor to detect shocked regions effectively. At reattachment, streamwise streaks form. The origin of these streaks is associated with either the G\"{o}rtler instability \cite{cao2019gortler} or baroclinic effects \cite{dwivedi2019reattachment} at reattachment. As in Case \ref{case:sbli}, the streaks correspond with a region of intense wall heating, which is shown in the Stanton number contour. Boundary layer transition occurs shortly after their formation and large turbulent structures are formed moving up the ramp. A close-up perspective is provided in Fig. \ref{fig:ramp/rampQualitativePicCloseUp3Annotated}. The streaks can be viewed conveniently using Q-criterion iso-surfaces. The large near-wall temperature gradient is evident at and after reattachment: the structures most upstream display the hottest temperatures and remain so near the wall. However, heat is dissipated moving in the wall-normal direction to the freestream and downstream of transition, where peak heat flux occurs. The Q-criterion structures also may show the second-mode instability, which is well-known to exist in hypersonic boundary layers and in ramp or flare configurations \cite{balakumar2005stability,benitez2023measurements}. In addition, secondary instabilities are evident, which are qualitatively similar to those in Cao et al. \cite{cao2019gortler}.

Fig. \ref{fig:ramp/afterReview_rampCpStProfile} shows the time and spanwise averaged Stanton number and pressure coefficient compared with the data from Cao et al. \cite{cao2022transition} and Roghelia et al. \cite{roghelia2017experimentala}. \textcolor{black}{We also considered two different grid sizes to serve as a grid-refinement study. For this study, we kept the number of cells in the $y$-direction constant, and divided/multiplied the number of cells in the $x$- and $z$-directions by a factor of 1.5 (see Table \ref{tab:rampGridDetails}). This resulted in a coarser grid of 15M cells and a finer grid of 75M cells.} The pressure coefficient was computed as:

\begin{equation}
    \mathrm{C_p} = \frac{2 \left( \left. p \right|_w - p_{\infty} \right) }{\rho_{\infty} u^2_{\infty}},
\end{equation}

\noindent Observing the pressure coefficient, separation is brought slightly upstream, however overall, there is good agreement with DNS and experiment. In the Stanton number plot, it is evident that the separation bubble behavior is well-captured, which is uncommon for equilibrium WMLES on account of the large non-equilibrium effects in this region. After reattachment, there is an over-prediction of wall heating; however, slightly farther downstream, the present WMLES matches the experimental measurements excellently. Previous WMLES studies of hypersonic compression ramps (\cite{ganju2021progress,van2022immersed,mettu2022wall}, have had difficulties matching DNS and/or experiment in the separation bubble region (and after). Indeed, Dawson et al. \cite{dawson2013assessment} showed that the inclusion of the pressure gradient term and \textcolor{black}{its} balancing convection term in a wall model are key to the accurate representation of the adverse pressure gradient region in compression ramp flows. Thus, the reduction of error here is encouraging. \textcolor{black}{With regards to the grid refinement study, all grid sizes seem to perform similarly. As above-mentioned, equilibrium WMLES does not perform well in regions of non-equilibrium flow. Therefore, we do see the separation point moving slightly upstream with grid refinement. Besides for this discrepancy, the differences between the grid sizes are minimal.}

\section{Conclusion} \label{sec:conclusion}

In this work, we proposed a low-dissipation GBR scheme to spatially discretize the convective fluxes in the compressible Navier-Stokes equations. The method takes advantage of characteristic transformation, which allows for selective treatment of the characteristic waves in the compressible Euler equations. A central scheme is achieved by averaging the left- and right-biased upwinded interpolations. This averaging is performed for all but the characteristic acoustic waves, which still require upwinded interpolations to maintain stability. The method was shown to be an effective and novel approach to minimizing unphysical dissipation. We also sought to observe differences using either primitive or conservative variables. 

The method was tested for benchmark problems, such as the two-dimensional shock entropy problem, two-dimensional viscous shock tube at $\mathrm{Re} = 2500$, and the three-dimensional inviscid Taylor-Green vortex. The method was shown to capture high-frequency peaks without inducing spurious oscillations in the shock entropy problem, match reference wall density more accurately in the viscous shock tube test case, and attain greater volume-averaged enstrophy for the inviscid Taylor-Green vortex case. We then tested the method in the context of WMLES, in which significant dissipation is already present due to extremely coarse computational grids. Two cases were considered: oblique shock impingement on a Mach 6 disturbed boundary layer, and Mach 7.7 transition to turbulence over a $15^{\circ}$ compression ramp. The results of the first test case displayed two important findings: a) centralizing the interpolation provided for full transition to turbulence and a good match with experimental data, and b) using conservative variables for characteristic transformation was very important. For the second, more challenging WMLES case, only MEG-C-CONS was considered. The time and spanwise averaged pressure coefficient and Stanton number were used as quantitative metrics to compare against previous experiment and DNS of the same case. The pressure coefficient profile showed an overall good match with experiment and DNS, although boundary layer separation was brought slightly upstream. The Stanton number profile showed satisfactory overall match, however, there was a large over-prediction in wall heating at reattachment and slightly downstream of reattachment. Notably, the separation bubble behavior as represented by the Stanton number was well-captured, which has been challenging for previous WMLES of compression ramp cases. Furthermore, after the region of wall heating over-prediction, the Stanton number matched excellently with the reference experimental profile. Errors in these quantitative metrics are very likely due to the equilibrium-assuming wall model, rather than the presented numerical method. As previously cited, the adverse pressure gradient caused by the compression ramp violates the convection-pressure gradient balance assumption of the equilibrium wall model.  

Future work will focus on applying the present approach to more complex flow scenarios, such as in WMLES of hypersonic cone-cylinder-flare configurations. Moreover, a challenging issue at hand is the use of a static cutoff value for the Ducros shock sensor. While the Ducros shock sensor is very effective at sensing shocks, the use of a non-universal, static cutoff parameter is a drawback. Ideally, the cutoff should be spatially varying and dependent on relevant flow features. As such, it is important to address these issues, especially in the context of hypersonic boundary layer transition.

\par\noindent\rule{\textwidth}{0.5pt}

\section*{Appendix}
\renewcommand{\thesubsection}{\Alph{subsection}}

\subsection{\textcolor{black}{Effect of choice of reconstruction variables on flow simulations}} \label{sec-appb}

\textcolor{black}{The proposed adaptive centralized scheme demonstrated different results when transforming to characteristic space using primitive or conservative variables. It was noted in Remark \ref{remark:which-variable} and in Chamarthi \cite{chamarthi2023efficient} that the flow structures could be different. In this Appendix, we carry out simulations for simple test cases using both primitive and conservative variables for characteristic transformation using the proposed schemes, MEG-C-PRIM and MEG-C-CONS, along with the TENO5 scheme \cite{fu2017targeted}. First, the order of accuracies of the considered methods are analyzed. Eqn. \ref{eqn:accu-euler} represents the initial profile that is convected in a computational domain of $[x, y] \in [-1, 1]$. The timestep is a function of the CFL and grid size: $\Delta t = \text{CFL} \Delta x^{2}$. The initial profile is convected until $t = 2$.} 

\begin{align}
    \left[ \rho, u, v, p \right] = \left[ 1 + 0.5 \sin (x+y), 1.0, 1.0, 1.0 \right]
    \label{eqn:accu-euler}
\end{align}

\textcolor{black}{Table \ref{tab:accu} presents the order of accuracies obtained for the considered schemes by computing the $L_2$ norm of the error between the exact and computed solution. The results demonstrate that the proposed schemes are fourth-order accurate in space for both primitive and conservative variables, consistent with Refs. \cite{chamarthi2023gradient,chamarthi2023implicit,chamarthi2023wave,chandravamsi2023application,chamarthi2023efficient}. Additionally, while the TENO5 scheme does indeed show fifth-order accuracy, the absolute error of the proposed schemes compared with TENO5, on both coarse and fine grids, is comparable. As noted in Ref. \cite{chamarthi2023gradient}, order of accuracy alone does not determine the superiority of a scheme. One of the advantages of GBR schemes is their superior dispersion and dissipation properties, which in the context of this work, allowed for the simulation of transition to turbulence on coarse grids.}

% Table generated by Excel2LaTeX from sheet 'Sheet1'
\begin{table}[H]
  \centering
  \footnotesize
  \caption{\textcolor{black}{Predicted order of accuracy for the considered schemes.}}
    \begin{tabular}{ c | cccccc }
    \hline
    \hline
        N   & TENO5 & Order & MEG-C-CONS  & Order & MEG-C-PRIM  & Order \\
        \hline
        $10^2$ & 6.79E-03 & -     & 9.58E-04 & -     & 9.50E-04 & - \\
        \hline
        $20^2$ & 2.24E-04 & 4.92  & 6.02E-05 & 3.99  & 6.02E-05 & 3.99 \\
        \hline
        $40^2$ & 7.06E-06 & 4.98  & 3.75E-06 & 4.00  & 3.75E-06 & 4.00 \\
        \hline
        $80^2$ & 2.21E-07 & 5.00  & 2.35E-07 & 4.00  & 2.35E-07 & 4.00 \\
        \hline
        \hline 
    \end{tabular}
    \label{tab:accu}%
\end{table}%

\textcolor{black}{In this second test case, we demonstrate the impact of characteristic transformation using primitive and conservative variables for MEG-C-PRIM, MEG-C-CONS, and the TENO5 scheme \cite{fu2017targeted}. The test involves two initially parallel shear layers that develop into two significant vortices at $t = 1$. All tests were run with a grid size of $N_x \times N_y = 96 \times 96$. The non-dimensional parameters for this test case are presented in Table \ref{tab:shearLayerNondimensionalParameters}}.

\begin{table}[h!]
    \centering
    \caption{\textcolor{black}{Parameters of the periodic double shear layer test case.}}
    \begin{tabular}{c c c c}
        \hline
        \hline
        $\mathrm{Ma}_{\infty}$ & $\mathrm{Re}_{\infty}$ & Pr & $\gamma$ \\
        \hline
        0.1 & 10,000 & 0.72 & 1.4 \\
        \hline
        \hline
    \end{tabular}
    \label{tab:shearLayerNondimensionalParameters}
\end{table}

\noindent \textcolor{black}{The initial conditions were:}

\begin{subequations}
   \textcolor{black}{ \begin{align}
        \rho &= \frac{1}{\gamma \mathrm{Ma}^2_{\infty}}, 
        \\[10pt]
        u &= 
        \begin{cases}
            \tanh \left[ 80 (y-0.25) \right], & \text{ if } (y \leq 0.5), \\
            \tanh \left[ 80 (0.75-y) \right], & \text{ if } (y > 0.5),
        \end{cases} 
        \\[10pt]
        v &= 0.05 \sin \left[ 2 \pi(x+0.25) \right], 
        \\[10pt]
        T &= 1.
    \end{align}}
\end{subequations}

\textcolor{black}{The reference solution, shown in Fig. \ref{fig:fin-dsl}, was computed with MEG-C-CONS on a grid of $N_x \times N_y = 512 \times 512$. For this test case, if the grid is under-resolved, unphysical braid vortices and oscillations can occur on the shear layers. Observing the well-resolved reference solution, there are no braid vortices or oscillations that form on the shear layers. Fig. \ref{fig:dpsl_72} displays the $z$-vorticity computed for MEG-C-CONS, MEG-C-PRIM, TENO5-CONS, and TENO5-PRIM. It is clear that MEG-C-CONS best approximates the reference solution, while the remaining methods result in unphysical braid vortices and oscillations. These findings suggest that numerical simulations can result in largely different flows depending on the choice of reconstructed variables for the cell interfaces. It is evident that the observed phenomenon is not specific to the MEG-C scheme but rather a general phenomenon that falls outside the scope of this paper.}

\begin{figure}[H]
    \centering
    \includegraphics[width=0.5\textwidth]{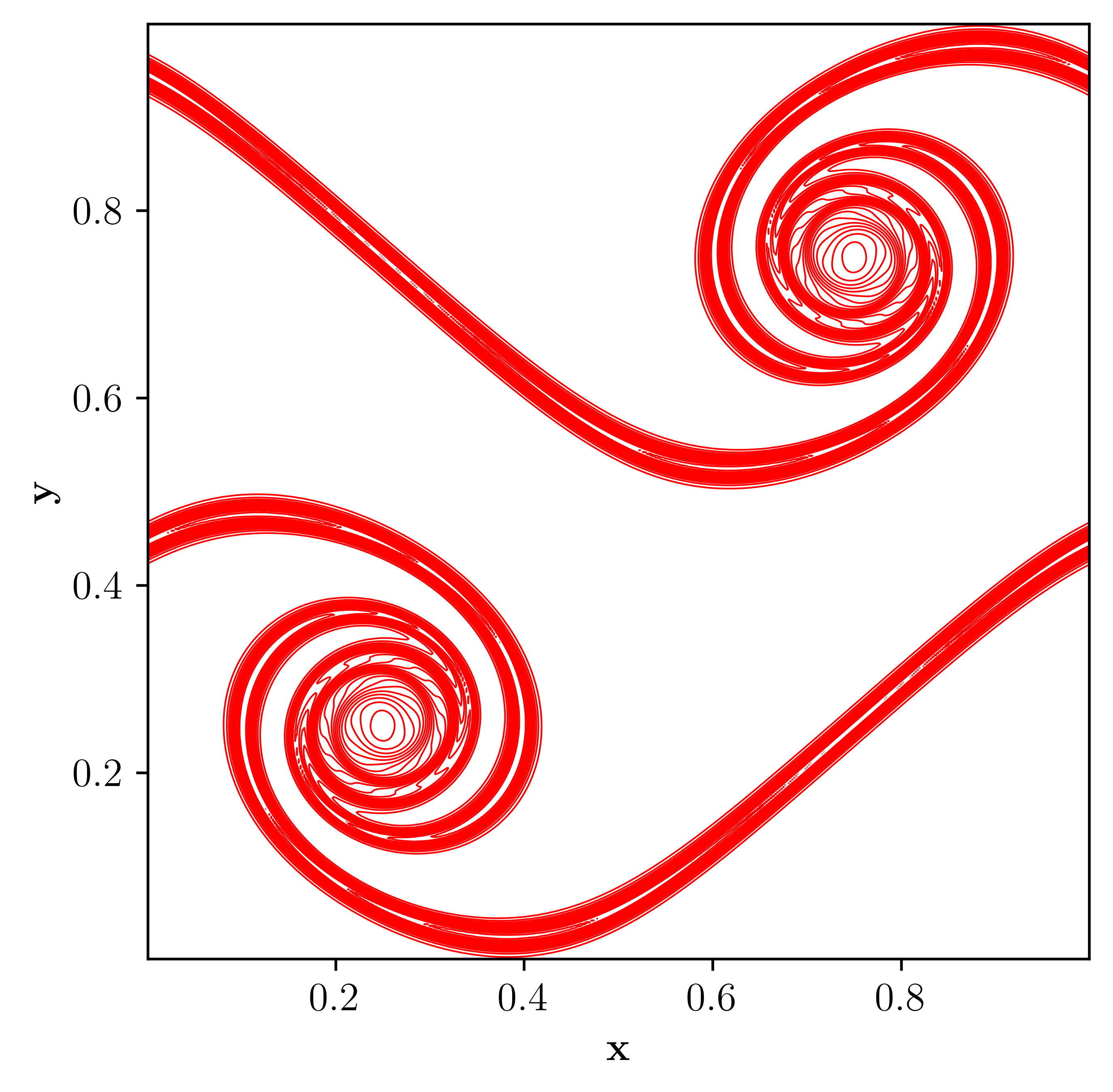}
    \caption{Reference $z$-vorticity contour computed on a grid of $512^2$.}
    \label{fig:fin-dsl}
\end{figure}

\begin{figure}[h!]
    \centering
    \begin{subfigure}{0.48\textwidth}
        \centering
        \includegraphics[width=\textwidth]{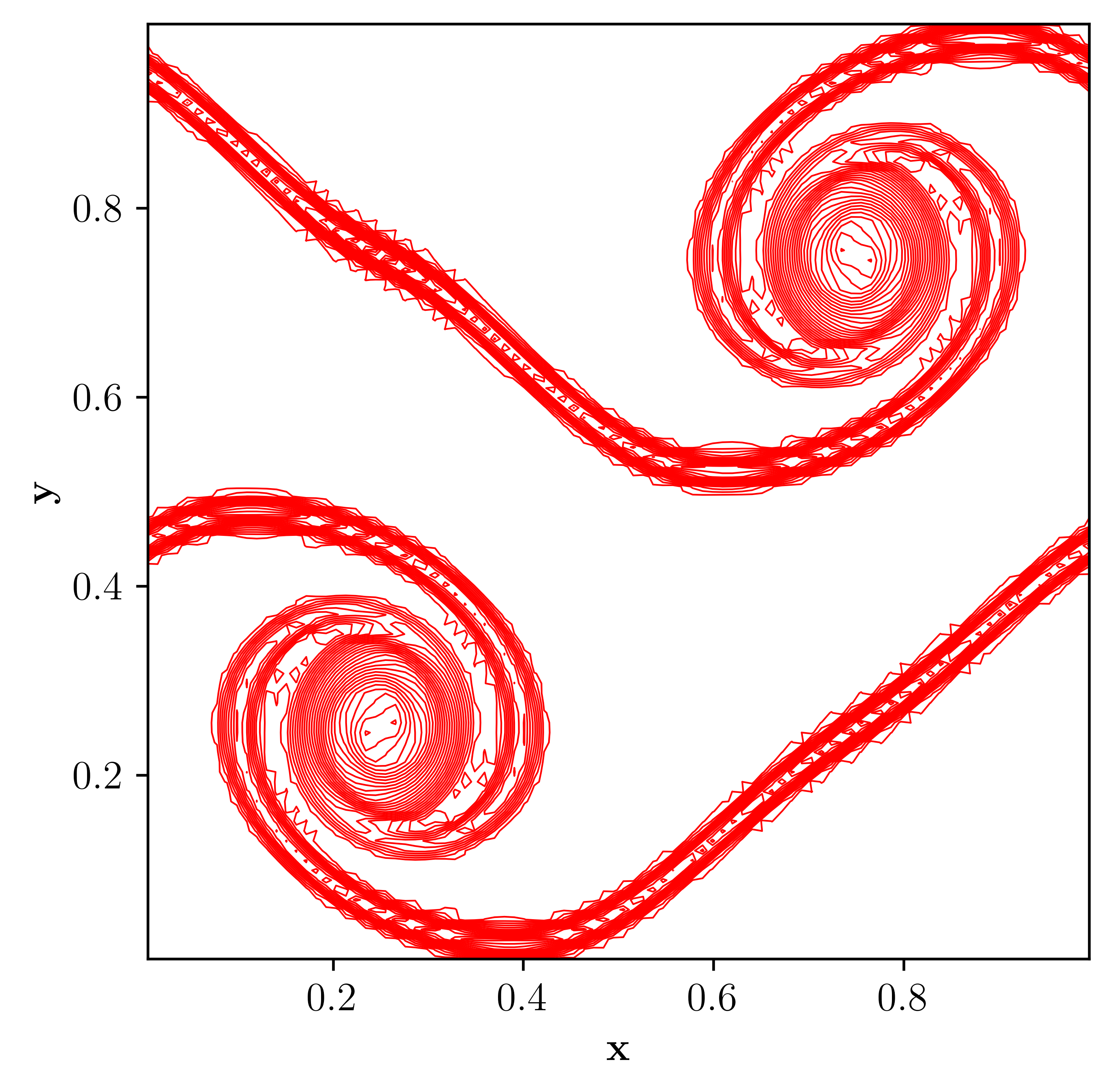}
        \caption{MEG-C-CONS.}
        \label{fig:meg-cons-pr}
    \end{subfigure}
    \hfill
    \begin{subfigure}{0.48\textwidth}
        \centering
        \includegraphics[width=\textwidth]{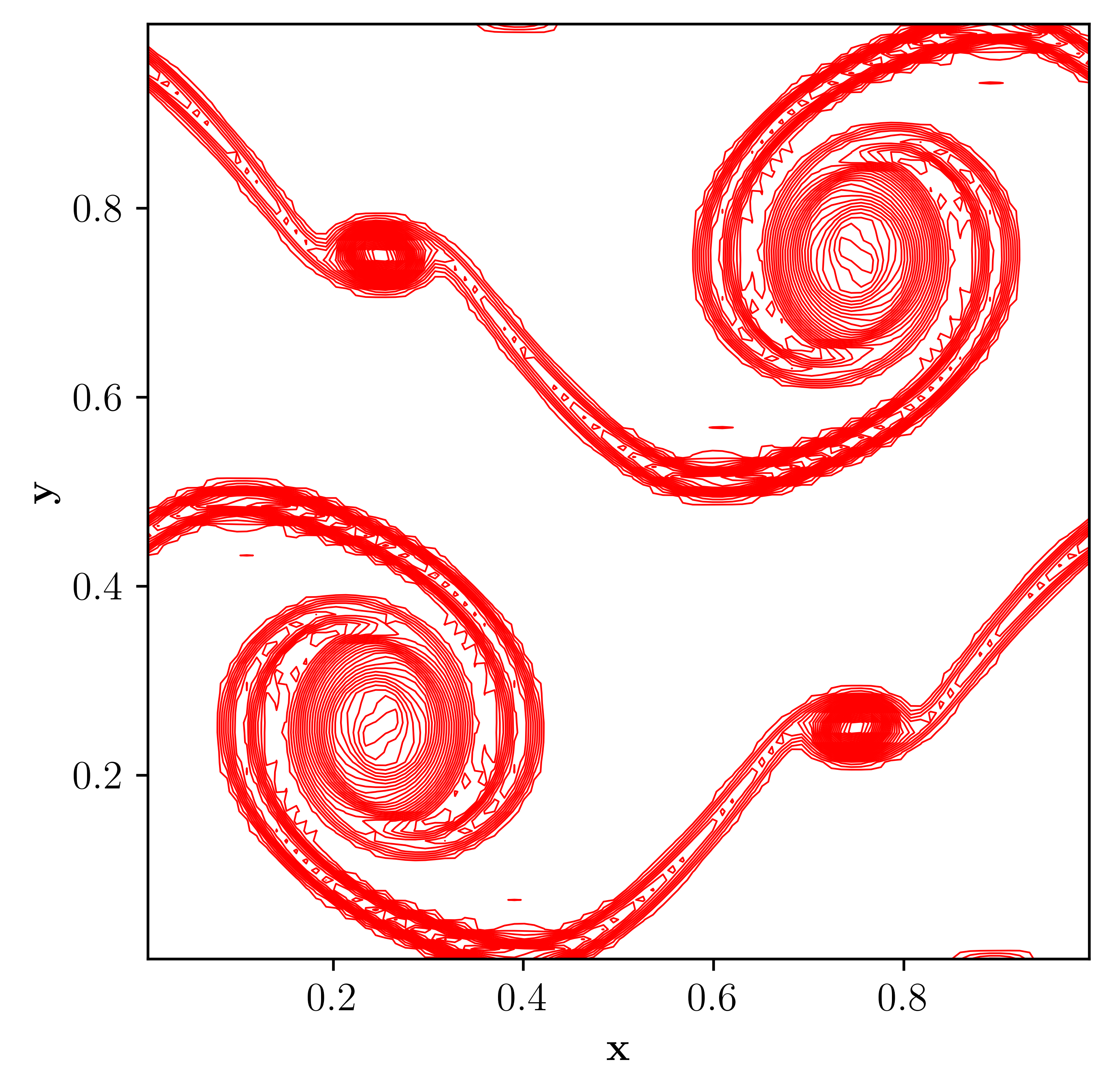}
        \caption{MEG-C-PRIM.}
        \label{fig:meg-prim}
    \end{subfigure}
    \vfill
    \begin{subfigure}{0.48\textwidth}
        \centering
        \includegraphics[width=\textwidth]{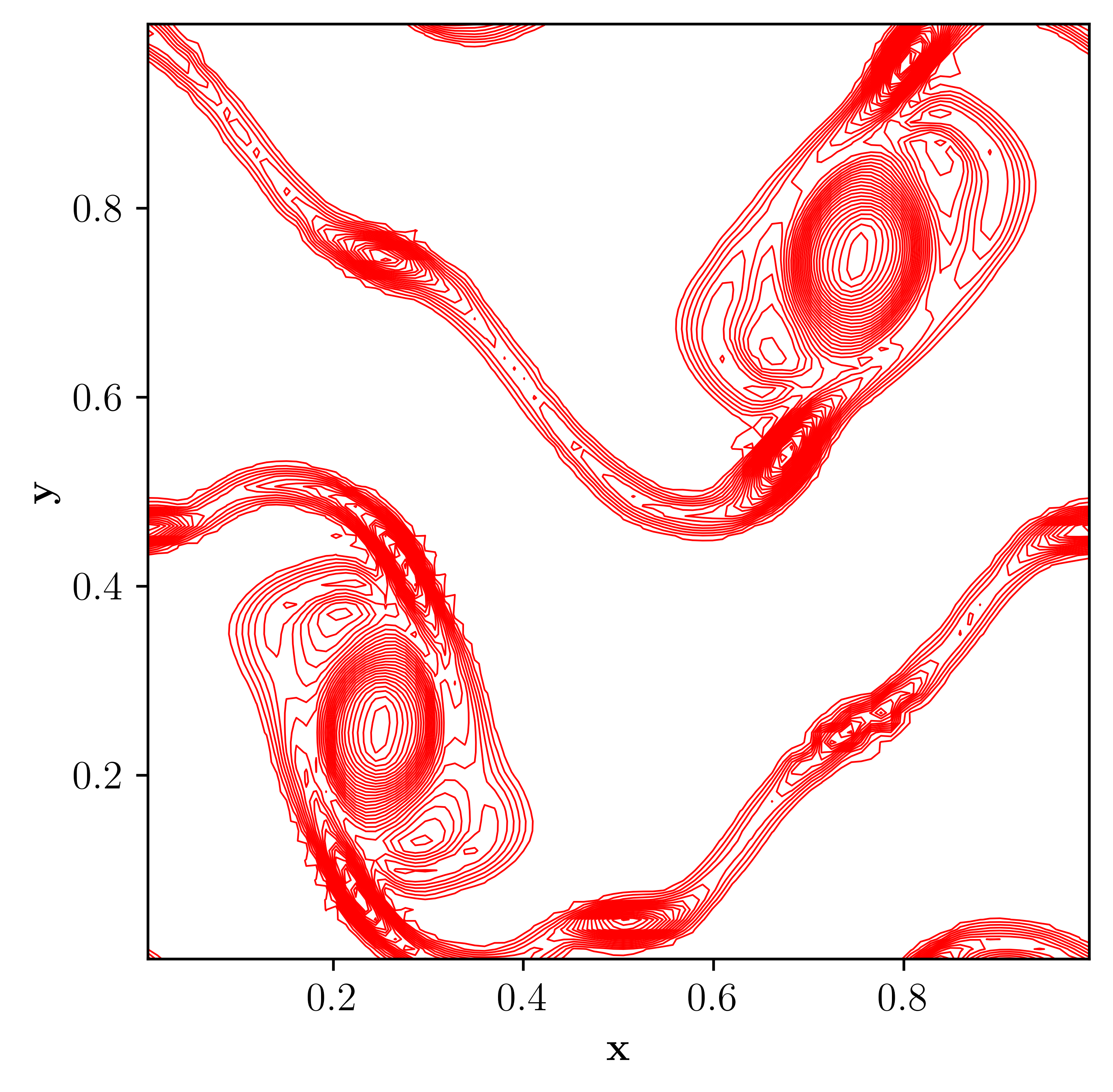}
        \caption{TENO5-CONS.}
        \label{fig:teno-cons}
    \end{subfigure}
    \hfill
    \begin{subfigure}{0.48\textwidth}
        \centering
        \includegraphics[width=\textwidth]{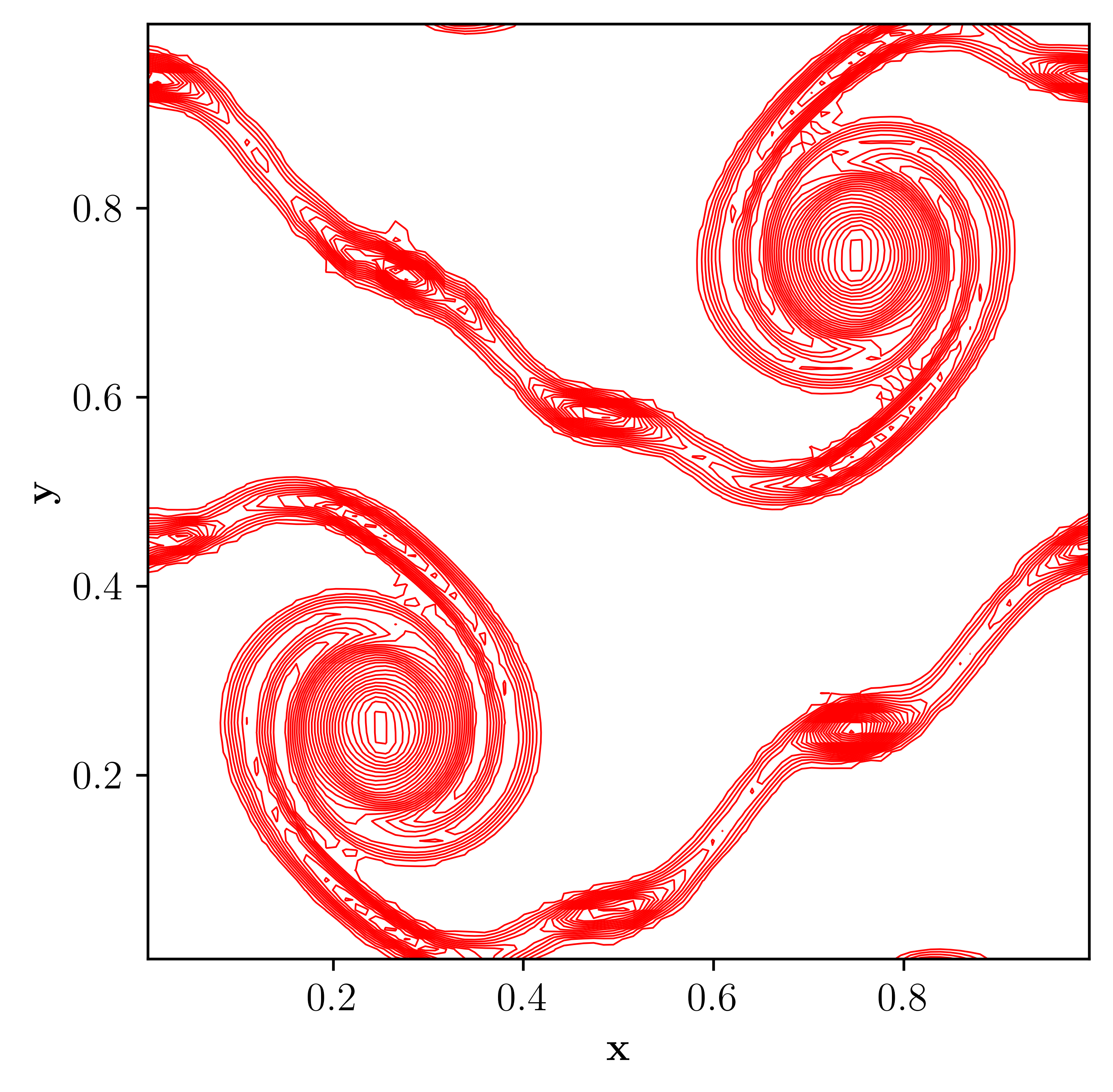}
        \caption{TENO5-PRIM.}
        \label{fig:ten-prim}
    \end{subfigure}
    \caption{$z$-vorticity contours of the considered schemes using a grid size of $96^2$.}
    \label{fig:dpsl_72}
\end{figure}

\subsection{\textcolor{black}{Computational Cost Analysis}}

\begin{table}[h!]
    \centering
    \caption{\textcolor{black}{Comparison of computational costs for the evaluated schemes using Case \ref{case:sbli}. The table lists times recorded over 100 iterations of the solver.}}
    \begin{tabular}{c c c c}
        \hline
        \hline
        MEG-S-PRIM & MEG-S-CONS & MEG-C-PRIM & MEG-C-CONS \\
        \hline
        10.59 s & 11.31 s & 10.71 s & 11.23 s \\
        \hline
        \hline
    \end{tabular}
    \label{tab:computationalCost}
\end{table}

\textcolor{black}{In this Appendix, we compare the considered schemes computational costs. We used Case \ref{case:sbli} without the wall model for the comparison. The recorded times are for 100 iterations of the solver. Observing Table \ref{tab:computationalCost}, the computational costs for all schemes are very similar; however, the schemes using conservative variables for characteristic transformation were slightly more computationally expensive. The increased computational cost is because of the computation of the conservative variable gradients. In general, the solver must compute the primitive variable gradients, no matter the convective scheme, because of the viscous flux spatial discretization scheme. However, the MEG-S-CONS and MEG-C-CONS schemes come with the extra associated cost of computing the gradient of the conservative variables.}

%% The Appendices part is started with the command \appendix;
%% appendix sections are then done as normal sections
%% \appendix

%% \section{}
%% \label{}

%% For citations use: 
%%       \citet{<label>} ==> Jones et al. [21]
%%       \citep{<label>} ==> [21]
%%

%% If you have bibdatabase file and want bibtex to generate the
%% bibitems, please use
%%
\bibliographystyle{elsarticle-num-names} 
\bibliography{sample.bib}

%% else use the following coding to input the bibitems directly in the
%% TeX file.

\end{document}